\documentclass[fleqn,usenatbib]{mnras}

\usepackage{newtxtext,newtxmath}

\usepackage[T1]{fontenc}

\DeclareRobustCommand{\VAN}[3]{#2}
\let\VANthebibliography\thebibliography
\def\thebibliography{\DeclareRobustCommand{\VAN}[3]{##3}\VANthebibliography}

\usepackage{graphicx}	
\graphicspath{{./}{figures/}}
\usepackage{amsmath}	
\usepackage{ulem}

\newcommand{\nuc}[2]{\ensuremath{^{#1}}{#2}}
\def\gsim{\lower.5ex\hbox{$\; \buildrel > \over \sim \;$}}
\def\lsim{\lower.5ex\hbox{$\; \buildrel < \over \sim \;$}}

\title[Origin of $^{60}$Fe nuclei in cosmic rays]{Origin of \nuc{60}{Fe} nuclei in cosmic rays: the contribution of local OB associations}

\author[N. de S\'er\'eville et al.]{
Nicolas de S\'er\'eville,$^1$\thanks{E-mail: nicolas.de-sereville@ijclab.in2p3.fr}
Vincent Tatischeff,$^1$
Pierre Cristofari,$^{1,2}$
Stefano Gabici$^3$
and Roland Diehl$^{4,5}$
\\
$^1$Universit\'{e} Paris-Saclay, CNRS/IN2P3, IJCLab, 91405 Orsay, France\\
$^2$Laboratoire Univers et Théories, Université de Paris, Observatoire de Paris, Université PSL, CNRS, 92190 Meudon, France\\
$^3$Universit\'e de Paris, CNRS, Astroparticule et Cosmologie, F-75006 Paris, France\\
$^4$Max Planck Institut f\"ur extraterrestrische Physik, Giessenbachstr.1, D-85748 Garching, Germany\\
$^5$Excellence Cluster Origins, Boltzmannstr. 2, D-85748 Garching, Germany\\
}

\date{Accepted XXX. Received YYY; in original form ZZZ}

\pubyear{2023}

\begin{document}
\label{firstpage}
\pagerange{\pageref{firstpage}--\pageref{lastpage}}
\maketitle

\begin{abstract}
The presence of live \nuc{60}{Fe} nuclei (lifetime of 3.8~Myr) in cosmic rays detected by the ACE/CRIS instrument suggests a nearby nucleosynthesis source. \nuc{60}{Fe} is primarily produced in core-collapse supernovae, and we aim to clarify whether the detected \nuc{60}{Fe} nuclei can be associated with a particular local supernova. We consider 25 OB associations and sub-groups located within 1 kpc of the solar system based on recent \textit{Gaia} census. A model is developed that combines stellar population synthesis within these OB associations, cosmic-ray acceleration within associated superbubbles, and cosmic-ray transport to the solar system. The most critical model parameter impacting \nuc{60}{Fe} cosmic-ray production is the explodability criterion, which determines if a massive star ends its life as a supernova. Our study points to the Sco-Cen OB association as the most probable origin of the observed \nuc{60}{Fe} nuclei, particularly suggesting they were accelerated in the Sco-Cen superbubble by a young supernova aged $\leq500$~kyr with a progenitor mass of approximately $13-20~M_\odot$. A less likely source is the supernova at the origin of the Geminga pulsar 342 kyr ago, if the progenitor originated in the Orion OB1 association. The contribution of local OB associations to the cosmic-ray density of stable \nuc{56}{Fe} is estimated to be around 20\%, with some sensitivity to cosmic ray acceleration efficiency and diffusion coefficient. These findings shed light on the origins of cosmic-ray nuclei, connecting them to nucleosynthesis events within our local cosmic neighborhood.

\end{abstract}


\section{Introduction} \label{sec:intro}
Cosmic rays (CRs) are believed to be a common component of the interstellar medium in galaxies, with an energy density that is comparable to the energy densities of other interstellar medium (ISM) components, such as the kinetic energy of bulk atomic or molecular gas motions, the thermal energy of hot plasma, and the magnetic energy of regular and turbulent fields \citep{Blasi:2013,gabici2019}. The consensual picture is that strong shock waves (of Mach number~$\gg$~$1$) accelerate CRs through diffusive shock acceleration (DSA)~\citep{Blandford:1978,Berezhko:1999,Lee:2012}. Such shock waves have typically been associated to massive star winds, supernovae and/or their remnants \citep{drury2012,Blasi:2013}. But several questions about CR acceleration remain poorly understood, including the spectra and maximum energy achieved by the CR particles at their sources and the efficiency of the acceleration process \citep{gabici2019}.
 
The propagation of CRs in the Galaxy after escaping from their sources also remains an important topic of research. The mean CR lifetime in the Milky Way ($\tau_{\rm CR} \sim 15$~Myr) is much longer than the light-crossing time ($< 0.1$~Myr), which is explained by diffusive confinement of the non-thermal particles by scattering on small-scale electromagnetic turbulence. Both pre-existing magnetohydrodynamic (MHD) turbulence \citep{Lazarian:2021,Lazarian:2023} and plasma waves self-generated by the CR streaming instability \citep{Kulsrud:1969,farmer:2004} are considered as scattering centers, but their relative importance for CR transport strongly depends on local plasma conditions in the multiphase ISM \citep{Kempski:2022}. Effective diffusion models are commonly used to describe CR propagation \citep[e.g.][]{evoli2019}, but the diffusion coefficient is hard to determine from first principles and may significantly vary within the Galaxy.

Recent gamma-ray observations of CR interactions with interstellar matter report significant variations of CR densities in specific regions, such as the Central Molecular Zone \citep{Hess:2016}, the inner Galaxy region between 1.5 and 4.5~kpc from the Galactic center \citep{Peron2021}, and the Cygnus region at a distance of 2-3 kpc from Earth (\citealp{Ackermann:2011,Astiasarain:2023}; see also discussions in \citealp{Aharonian:2019} and \citealp{Bykov:2022}). Significant variations of the measured CR-induced ionisation rate in molecular clouds also point to variation of the density of low-energy CRs throughout the Galaxy \citep{Indriolo:2012,gabici2022,Phan:2023}. In particular, the local spectrum of MeV CRs measured by the Voyager probes may not be representative of the low-energy CR spectrum elsewhere in the Galaxy \citep{phan2021}. In addition, according to \cite{Kachelriess2018}, the unexpected hardness of the CR positron and antiproton spectra above $\sim 100$~GeV can be explained by a significant contribution to the CR flux of particles accelerated in a local supernova some 2-3 Myr ago.

The detection of \nuc{60}{Fe} nuclei in CRs  with CRIS on the ACE spacecraft \citep{binns2016} offers a unique opportunity to study the contribution of localized and nearby sources to the CR population seen here, hence addressing CR source and transport simultaneously. $^{60}$Fe is a primary CR, i.e. it is not produced to any significant extent by nuclear spallation of heavier CRs in the ISM. It is thought to be synthesized mainly in core-collapse supernovae of massive stars. Its radioactive lifetime of 3.8~Myr is sufficiently long such it can potentially survive the time interval between nucleosynthesis and detection at Earth. But the $^{60}$Fe lifetime is significantly shorter than $\tau_{\rm CR}$, which suggests that nucleosynthesis sites far out in the Galaxy are plausibly beyond reach for $^{60}$Fe CRs surviving such a journey.  

$^{60}$Fe has also been found in sediments from the Pacific oceanfloor~\citep{knie2004}, complemented by findings in other sediments across Earth and even on the Moon \citep{Wallner:2016,Wallner:2021}. Its live presence on Earth, combined with its radioactive decay time, and with typical velocities for the transport of interstellar matter (transport of $^{60}$Fe to Earth generally assumes adsorption on dust grains travelling at velocities of the order of $\sim 10$~km~s$^{-1}$), suggested that it may be due to recent nucleosynthesis activity near the solar system. 

In parallel to the CR measurements, and to the recent data obtained on $^{60}$Fe in sediments and on the Moon, our knowledge of the distribution of stars, and especially massive stars and OB associations in our local environment within a few kiloparsec is rapidly increasing, as recently illustrated with $Gaia$ observations \citep{Zucker2022a,zucker2022}. In the problem of the origin of CRs, OB associations are especially relevant, since they are expected to substantially enrich the ISM, injecting nuclear material through their winds and when exploding. The potential important contribution of OB associations in the CR content has been discussed in several works~\citep{parizot2004,binns2007,murphy2016,tatischeff2021}. The recent $^{60}$Fe data and ever increasing knowledge on the local OB associations, provides an opportunity for probing the contribution of OB associations to CRs. 

In this paper, we aim to set up a bottom-up model for the origin of $^{60}$Fe in CRs near Earth, based on modelling  both the plausible nearby massive star groups as sources of the nucleosynthesis ejecta including $^{60}$Fe, together with modelling the acceleration near the sources, and the transport through the specifics of ISM trajectories from the sources to near-earth space.  We rely on Monte-Carlo simulations, developing a model combining a description of the OB stellar population, accounting for CR acceleration and transport, and confront it to available $^{60}$Fe data. The model also allows us to discuss the origin of other CR nuclei such as $^{56}$Fe and $^{26}$Al.

This paper is organised as follows. First we convert the measurement data of $^{60}$Fe in CRs into interstellar fluxes (Section~\ref{sec:cris}). Then  we present our population synthesis model for determination of time-dependent production of $^{60}$Fe, followed by CR acceleration and transport (Section~\ref{sec:model}). We apply this to nearby massive-star groups (Section~\ref{sec:OB}), and evaluate these results towards constraints for locally found $^{60}$Fe CRs (Section~\ref{sec:results}). We conclude with a discussion of the sensitivity of our findings to various assumptions and ingredients of this bottom-up modelling.

\section{Density of $^{60}$Fe and $^{26}$Al CRs from ACE/CRIS measurements} \label{sec:cris}
ACE/CRIS collected $^{56}$Fe and $^{60}$Fe CR nuclei between $\sim 195$ and $\sim 500$~MeV~nucleon$^{-1}$, reporting 15 $^{60}$Fe CR nuclei \citep{binns2016}. The reconstructed mean energy at the top of the CRIS instrument is 340~MeV~nucleon$^{-1}$ for $^{56}$Fe and 327~MeV~nucleon$^{-1}$ for $^{60}$Fe. According to \citet{binns2016}, the CR modulation inside the solar system during the 17-year period of the data taking can be accounted for with an average force-field potential $\phi = 453 $~MV, corresponding to an energy loss of 210~MeV~nucleon$^{-1}$ for $^{56}$Fe and 196~MeV~nucleon$^{-1}$ for $^{60}$Fe. Thus, the mean energies in the local interstellar space are 550~MeV~nucleon$^{-1}$ for $^{56}$Fe and 523~MeV~nucleon$^{-1}$ for $^{60}$Fe, and the corresponding velocities are $0.778c$ and $0.768c$ ($c$ is the speed of light).

The measured iron isotopic ratio near Earth is $(^{60}{\rm Fe}/^{56}{\rm Fe})_{\rm CRIS} = (4.6 \pm 1.7) \times 10^{-5}$ \citep{binns2016}. The flux ratio in the local ISM (LISM) can be estimated from the force-field approximation to the transport equation describing the CR modulation in the heliosphere \citep{gleeson1968}. In this simple model, the CR flux in the LISM is related to the one measured near Earth by a shift in particle momentum, which gives for the Fe isotopic ratio: 
\begin{eqnarray}
     (^{60}{\rm Fe}/^{56}{\rm Fe})_{\rm LISM} & = & (^{60}{\rm Fe}/^{56}{\rm Fe})_{\rm CRIS} \times (p_{60, {\rm LISM}} / p_{60, {\rm CRIS}})^2 \times \nonumber \\ 
     & & (p_{56, {\rm CRIS}} / p_{56, {\rm LISM}})^2,
 \end{eqnarray}
where $p_{56, {\rm CRIS}}=48.5$~GeV/$c$, $p_{60, {\rm CRIS}}=50.8$~GeV/$c$, $p_{56, {\rm LISM}}=64.5$~GeV/$c$ and $p_{60, {\rm LISM}}=67.0$~GeV/$c$ are the $^{56}$Fe and $^{60}$Fe mean momenta at the top of the CRIS instrument and in the local ISM. Thus, $(^{60}{\rm Fe}/^{56}{\rm Fe})_{\rm LISM} = (4.5 \pm 1.7) \times 10^{-5}$.

The spectrum of $^{56}$Fe CRs in the LISM can be estimated from the work of \citet{boschini2021}, who used recent AMS-02 results \citep{aguilar2021}, together with Voyager~1 and ACE/CRIS data, to study the origin of Fe in the CR population. Their calculations are based on the \textsc{GalProp} code to model the CR propagation in the ISM \citep{strong1998} and the \textsc{HelMod} model to describe the particle transport within the heliosphere \citep{boschini2019}. Integrating the iron spectrum given by these authors in the energy range from $400$--$700$~MeV~nucleon$^{-1}$, which approximately corresponds to the range of the CRIS measurements, we find $I_{\rm LISM}({\rm Fe}) = 2.9 \times 10^{-5}$~cm$^{-2}$~s$^{-1}$~sr$^{-1}$ and the density $n_{\rm LISM}({\rm Fe}) = I_{\rm LISM} \times 4 \pi / v = 1.6 \times 10^{-14}$~CR~cm$^{-3}$. The intensity of $^{60}$Fe CRs between $400$ and $700$~MeV~nucleon$^{-1}$ in the LISM is $I_{\rm LISM}({\rm ^{60}Fe}) = I_{\rm LISM}({\rm Fe}) \times (^{60}{\rm Fe}/^{56}{\rm Fe})_{\rm LISM} = (1.3 \pm 0.5) \times 10^{-9}$~cm$^{-2}$~s$^{-1}$~sr$^{-1}$ and the density in this energy range is $n_{\rm LISM}({\rm ^{60}Fe}) = (7.1 \pm 2.7) \times 10^{-19}$~CR~cm$^{-3}$.  

Recently, \citet{boschini2022} found that the aluminium CR spectrum measured by AMS-02 presents a significant excess in the rigidity range from $2$--$7$~GV compared to the spectrum predicted with the \textsc{GalProp}~--~\textsc{HelMod} framework from spallation of $^{28}{\rm Si}$ CRs and heavier nuclei. They suggested that this excess could be attributed to a source of primary CRs of radioactive $^{26}$Al (half-life $T_{1/2}=7.17\times10^5$~yr) possibly related to the well-known $^{22}$Ne excess in the CR composition. The latter is interpreted as arising from acceleration of massive star wind material in OB associations \cite[see][and references therein]{tatischeff2021}. Here, we study the contribution of primary $^{26}$Al CRs originating together with $^{60}$Fe from local OB associations. 

ACE/CRIS measured $(^{26}{\rm Al}/^{27}{\rm Al})_{\rm CRIS}=0.042 \pm 0.002$ between $125$ and $300$~MeV~nucleon$^{-1}$, corresponding to the LISM energy range $190$--$560$~MeV~nucleon$^{-1}$ \citep{yanasak2001}. From the Al spectrum in the LISM computed by \citet{boschini2022}, we find the mean energy of Al CRs in the LISM to be 355~MeV~nucleon$^{-1}$ ($v=0.690c$) and the LISM density of Al CRs between $190$ and $560$~MeV~nucleon$^{-1}$ to be $n_{\rm LISM}({\rm Al}) = 1.0 \times 10^{-14}$~CR~cm$^{-3}$. The $^{26}$Al CRs density in this energy range is then $n_{\rm LISM}({\rm ^{26}Al}) = (4.2 \pm 0.2) \times 10^{-16}$~CR~cm$^{-3}$.

\section{CR population synthesis and transport} \label{sec:model}
\begin{figure}
    \includegraphics[width=0.5\textwidth]{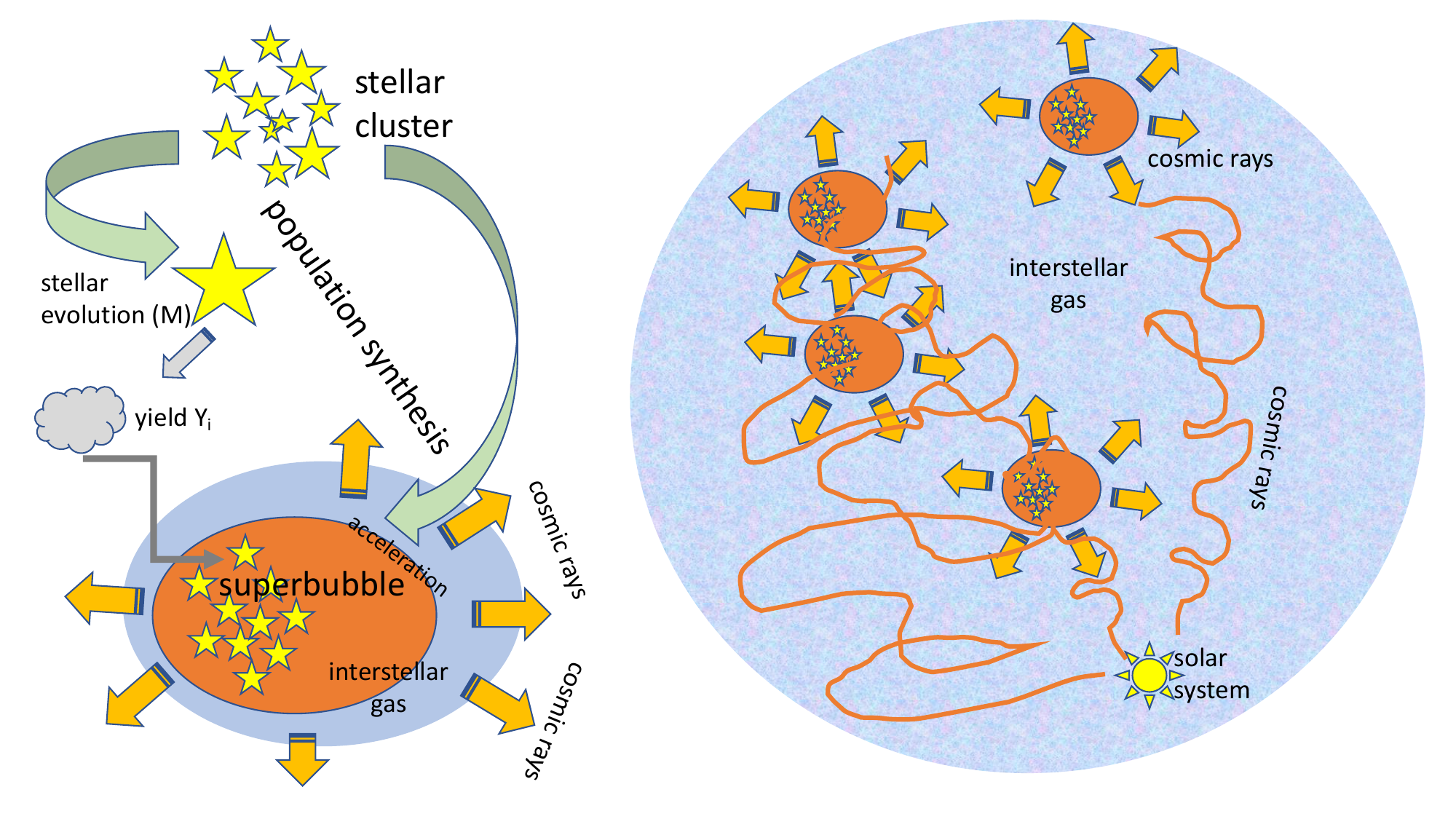}
    \caption{Illustration of our bottom-up CR source model.
    The population synthesis of a massive star cluster (left) evaluates
    stellar evolution and supernova models to determine the abundance
    of particular isotopes within the cluster gas and its superbubble.
    This gas is accelerated to CRs, which are transported through
    galactic ISM towards the Sun (right).}
    \label{f:modelSketch}
\end{figure}

We developed a bottom-up model (Fig.~\ref{f:modelSketch}) for the CR flux at the solar system, integrating contributions from the presumed sources of radioactive nuclei within massive-star clusters. Basic ingredients are the yields of ejecta from stars and supernovae. For each cluster, its age and richness are used together with a generic initial-mass distribution to determine proper weighting, thus building a time profile of interstellar nuclide abundances for and within each specific cluster. With plausible assumptions about CR acceleration efficiency within such a massive-star group and the likely superbubble configuration resulting from the clustered stellar and supernova activity, we derive a CR source density for each star cluster, as it varies with time. Propagation of these CRs towards the solar system requires a CR transport model that accounts for the location of the source within the Galaxy and its distance from the solar system, accounting for specifics of CR transport in the solar neighborhood. Integrating contributions of all sources from which CRs could have reached instruments near Earth in our present epoch, we thus obtain a bottom-up determination of the local CR flux in terms of model parameters based on stars and supernovae.

Our model is similar to and builds on those of \cite{Gounelle2009, Voss2009, Young2014} for example, and we focus on the activity of massive stars ($M \geq 8M_\odot$) in OB associations. The novelty of our work is to couple the nucleosynthesis output of a massive-star group to a CR transport model, which then allows the prediction of the flux of CRs near Earth. Adjusting parameters of our model to best match CR data taken near Earth, we can therefore constrain the origins of locally-observed CRs, back-tracing them to the contributing massive star groups.

\subsection{Radioisotope production at CR sources}
We aim to know the production of radioactive isotopes from the ensemble of stars and supernovae in the nearby Galaxy. \textit{Population synthesis} is the tool commonly used to predict the integrated outcome and properties of stellar populations \citep{Cervino2006,Cervino2013}. This approach has been used in particular to predict the radionuclide enrichment of the ISM near OB associations using such a bottom-up approach that implements our knowledge about star, their evolution, and their nucleosynthesis yields \citep{Voss2009}. In the following, we describe key aspects of the stellar population synthesis part of our model.

\subsubsection{Population synthesis ingredients}
\textbf{Initial mass function.} 
A population of stars that formed simultaneously and within the same environment, such as in a cluster, is characterised by the distribution in mass of the stars after having been formed, the \textit{initial mass function} (IMF). Observationally, the stellar population seen within a cluster reflects the \textit{current mass distribution}. From this, one may estimate an initial mass distribution by corrections for the stars of high mass that already may have disappeared, when the cluster age is known, or can reliably be estimated. There is considerable debate of how generic the initial mass distribution may be, or how it may depend on the feedback properties for different stellar density and interstellar gas density \citep[e.g.][]{Kroupa:2019}. But the widely observed similarity of the power-law shape of the mass distribution \citep{Kroupa2001} suggests that the mass distribution of newly-formed stars is a result of the physical processes during star formation, as it may be inhibited or modified by energetic feedback from the newly-formed stars. The IMF was initially described for intermediate to large stellar masses by a single power law function by~\cite{Salpeter1955}. Toward the low mass end down to the brown dwarf limit the IMF flattens and can be described by a log-normal shape \citep{Miller1979}, or a broken power-law \citep{Kroupa2001}. Our model implements any IMF described by a multi-part power law, and we use as default the parameters given by \cite[Eq. 6]{Kroupa2001}; this gives an average mass of the association members of 0.21~$M_{\odot}$ and a fraction of stars having a mass greater than 8~$M_{\odot}$ of $1.6\times10^{-3}$. The stellar content of specific known OB associations is mostly derived from a census of bright stars such as O and B stars ($M\geq2.8~M_\odot$ \citep{habets1981}), thus only the high-mass end of the IMF is relevant. The upper end of the mass distribution for massive stars is debated \citep[e.g.][]{Heger:2003,Vanbeveren:2009a,Schneider:2018}. Theoretical uncertainties derive from the star formation processes for very massive stars as nuclear burning sets in during the mass accretion phase, but also from late evolution of massive stars towards core collapse that may be inhibited by pair instability. Observationally, stars with masses up to 300~$M_\odot$ have been claimed to exist in the LMC's 30Dor region \citep{Schneider:2018}. In our model we consider an upper limit of $120~M_\odot$, which seems reasonable compared to the observational upper limit for single stars in our Galaxy of $\sim150~M_\odot$ \citep{Maiz2007}. This allows us to use the full range of the mass grid for stellar yields from \cite{LC18}.

\begin{figure}
    \includegraphics[width=0.5\textwidth, trim= 0 0 0 20, clip]{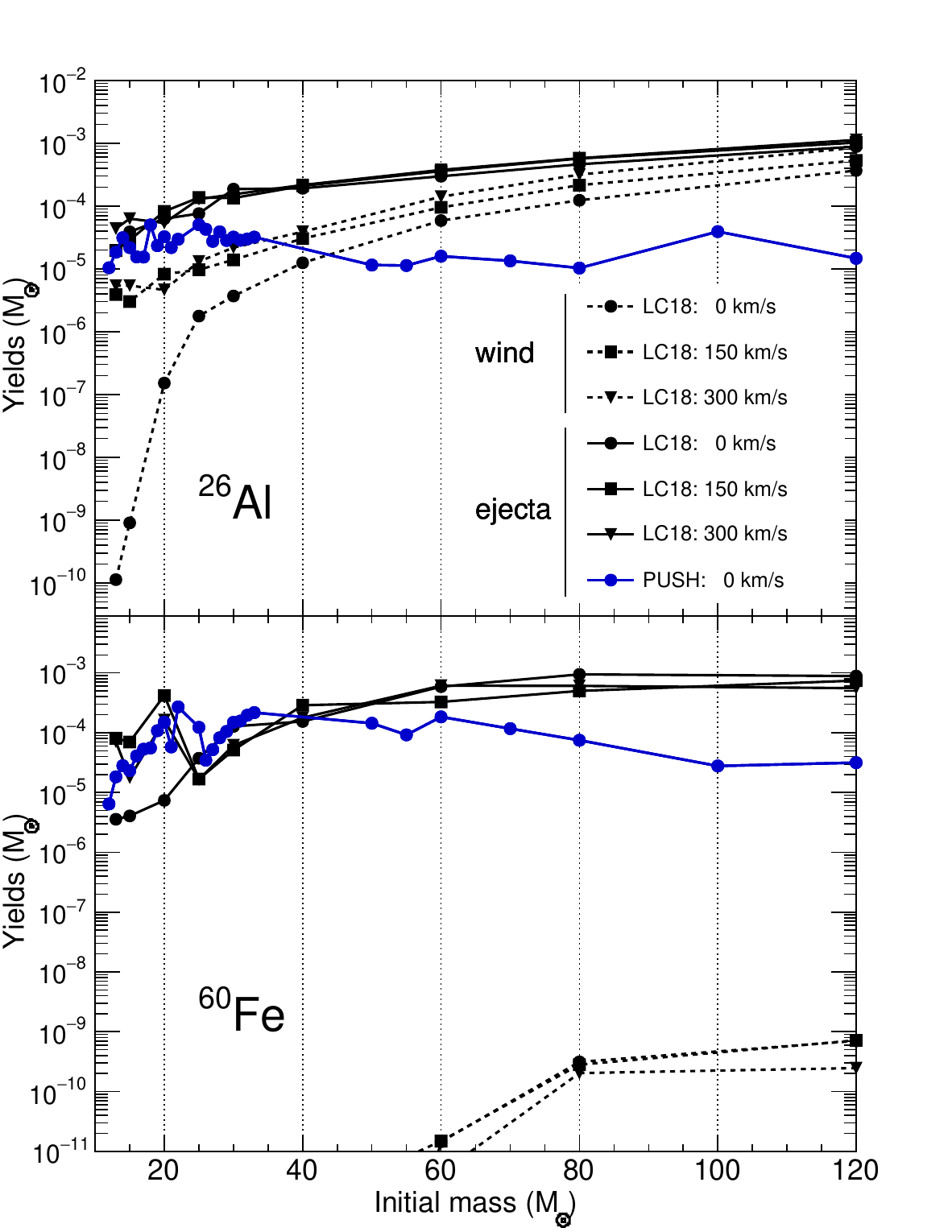}
    \caption{Wind (dashed lines) and explosion (solid lines) yields for $^{26}$Al (top) and $^{60}$Fe (bottom) nuclides produced by solar metallicity stars. Yields for stellar models from \protect\cite{LC18} (LC18; set M) and \protect\cite{ebinger2019} (PUSH; based on the pre-explosion models of \protect\citealp{WH07}) are displayed for non-rotating stars and rotating ones with an initial rotational velocity of 150 and 300~km s$^{-1}$.}
    \label{f:yields}
\end{figure}

\textbf{Stellar yields.} 
Massive stars contribute significantly to the enrichment of the ISM by releasing nuclear processed material through stellar winds and during their explosive phase. Models of stellar evolution that include a detailed nucleosynthesis network trace stars through all evolutionary stages and thus predict the nuclide yields both from the stellar winds and the supernova explosion phase. An example of \nuc{26}{Al} and \nuc{60}{Fe} yields as a function of the initial stellar mass is presented in Fig.~\ref{f:yields} for models from \cite{LC18} and \cite{ebinger2019}. Comparing the yields for non rotating stars gives an idea of the systematic uncertainties of these models. The contribution of explosive nucleosynthesis (solid lines) typically amounts to $\sim10^{-5}-10^{-3}~M_\odot$ of ejected mass for both nuclides, with a mild dependence to the initial stellar mass. The wind contribution is completely negligible for \nuc{60}{Fe}, quite in contrast with the case of \nuc{26}{Al} where this is very significant for the high-end massive stars, even comparable to the contribution from the explosion. In Fig.~\ref{f:yields}, yields are also given as a function of the star's initial rotational velocity. Massive stars are known to be rotating objects \citep{Glebocki2005}, and yields are affected through mixing processes stimulated by stellar rotation. Indeed, rotation induces a slow mixing of both fresh fuel from the envelope into the burning core and of freshly synthesized material from the burning H-core into the envelope of the star. Stellar rotation also enhances the ejection into the ISM with stronger winds \citep{Meynet2000}. This leads to a larger \nuc{26}{Al} wind contribution for stellar models that include such rotation (see Fig.~\ref{f:yields}). The large difference observed between rotating and non-rotating models from \cite{LC18} in the low-mass range ($\leq30~M_\odot$) is due to the treatment of rotational mixing and the impact of a dust driven wind on the stellar mass loss \citep{CL13}. The effect of rotation on the explosive yields is more difficult to assess, with no clear enhancement for rotating models except for the case of \nuc{60}{Fe} for stars up to $20~M_\odot$, where the yields are about 10 times larger than for the non-rotating models. 
In our population synthesis model we follow the prescription of \cite{Prantzos2018} where the initial distribution of rotational velocity of stars is constrained from the study of the production of $s$-elements with a galactic chemical evolution model. We therefore consider that the probability for solar metallicity OB stars to have an initial rotational velocity of 0-, 150- and 300-km s$^{-1}$ is 67\%, 32\% and 1\%, respectively.

\textbf{Stellar explodability.} The nucleosynthetic output from a massive star strongly depends on its fate during its gravitational collapse at the end of its evolution. Massive stars which collapse and form black holes, either directly or through fallback, are not expected to enrich the ISM, while their successful explosion will disseminate freshly synthesised nuclear material into the ISM. Which star of a specific mass may experience a successful explosion and for which stellar mass this fails is an actively debated question \citep[e.g.][]{Foglizzo2015,Sukhbold2016}. Even though there are several observations of ccSNe with indication of the progenitor mass \citep[][Appendix A]{ebinger2019}, the initial mass uncertainty and the low rate of ccSNe make it difficult to constrain the explodability from observations only. For simplicity, some models assume that massive stars collapse directly to black holes when their initial mass is greater than  $25~M_\odot$ \citep{LC18}, while others claim a transition mass in the range $100~M_\odot\leq M\leq140~M_\odot$ \citep{Janka2012}. Detailed numerical treatment of the explosion of massive stars has suggested that their explodability depend on the compactness in their pre-SN phase \citep{Oconnor2011}, which leads to irregular gaps within the range of the stellar initial masses where massive stars undergo a successful explosion \citep[e.g.][]{Sukhbold2016, ebinger2019}. 

\textbf{Fiducial model.} In the present work we use as nominal set of parameters the IMF from \cite{Kroupa2001} (see also \citealp{Kroupa2002}) with the stellar evolution prescription and yields from \cite{LC18}. For explodability we assume that only stars below $\leq25M_\odot$ explode as ccSN and subsequently release ejecta in the ISM. This corresponds to the case of set R defined in \cite{LC18}, which is equivalent to set M (displayed in Fig.~\ref{f:yields}) where the explosion yield is set to zero above $25M_\odot$. Both stellar yields and lifetime depend on the star metallicity. However, since most OB associations are relatively young with typical ages below 50~Myr~\citep{Wright2020} we adopt stellar yields and lifetimes for solar metallicity stars. Concerning the initial distribution of rotational velocities of stars we follow the prescription of \cite{Prantzos2018}. The flexibility of our model allows to switch for different IMFs, stellar and explosion yields, and explodability criteria very easily. We investigate the impact of changing these input parameters and describe this in Sec.~\ref{sec:parameters}.

\subsubsection{Nuclide enrichment of the gas in OB association}
As starting point of our population synthesis model we sample the IMF to generate the masses of the OB association members. We use the IMF function described in \cite{Kroupa2001,Kroupa2002}, and only massive stars ($M\geq8M_\odot$) are considered. We use random sampling for simplicity, considering the difference to optimal sampling \citep[see, e.g.][]{Yan:2023} rather insignificant for our purposes. The sampling procedure is repeated until a given total stellar content of the OB association is reproduced. This content can be deduced from the observations (see Sec.~\ref{sec:OB}) or it can be specified \textit{a priori} as a total number of stars integrated over the full IMF mass range. For each massive star, an initial rotational velocity is randomly generated. Then, the lifetime of the star, which depends on the initial stellar mass and rotational velocity, is determined from stellar evolution models, and their nucleosynthesis yields during evolution are assembled. The contribution of stellar winds of massive stars is also taken into account in our model: for simplicity, stellar winds are assumed to be released at the end of the star lifetime since we are mainly interested in \nuc{60}{Fe} which is not significantly produced by stellar winds. For massive stars ending their lives as ccSNe as controlled by the explodability criterion, their ejecta are released at their time of explosion.

It is usually assumed that massive stars within a stellar cluster can be considered as a coeval population \citep{Lada2005}. Thus, the temporal evolution of the mass of a radionuclide $M(j,t)$ in the gas of an OB association is calculated as the sum of the individual contributions associated to each massive star:
\begin{equation}
    M(j,t) = \sum_{i=1}^{n}\left[Y_i^{\rm wind}(j) + \eta_i\times Y_i^{\rm expl}(j)\right] e^{-(t-t_i)/\tau_{j}},
    \label{e:mass}
\end{equation}
where $Y_i^{wind}(j)$ and $Y_i^{expl}(j)$ are the wind and explosive yields, respectively, for nuclide $j$ associated to the $i^{\rm th}$ massive star with stellar lifetime $t_i$. $\eta_i$ is a parameter taking value of 0 or 1 whether the considered star explode as a ccSN or not depending on the adopted explodability criterion. The exponential term reflects the free radioactive decay of nuclide $j$ according to its corresponding lifetime $\tau_j$. This term should be set to 1 in case of stable nuclides.

\begin{figure}
    \begin{center}
        \includegraphics[width=0.5\textwidth]{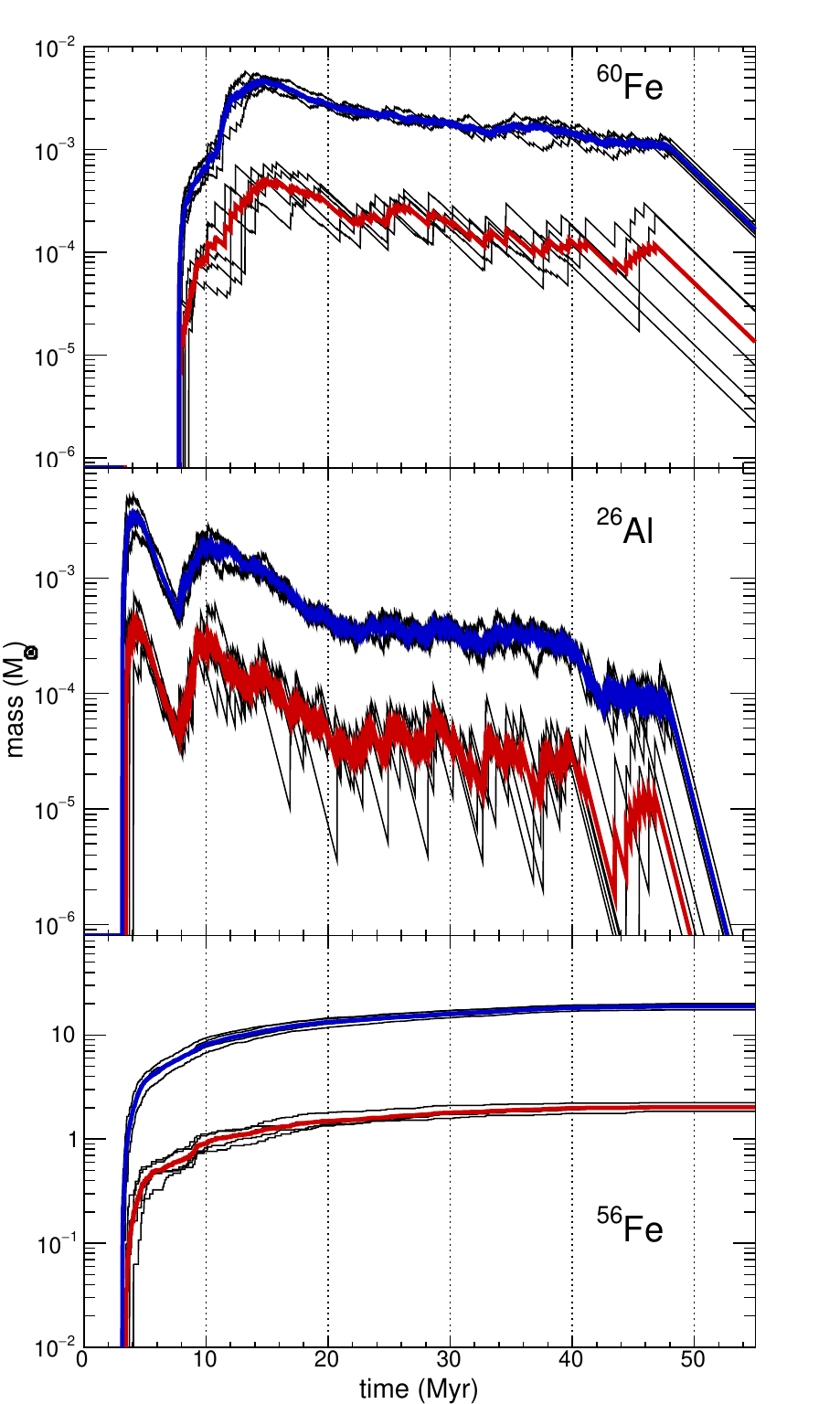}
        \caption{Time evolution of the mass of \nuc{60}{Fe}, \nuc{56}{Fe}, and \nuc{26}{Al} produced by an OB association with a stellar content of $10^4~M_\odot$ (red) and $10^5~M_\odot$ (blue). The time origin corresponds to the birth of the OB association which is evolved during 55~Myr. Five Monte Carlo realizations are represented in each case (black lines), and the associated average is shown by the colored solid line.}
        \label{f:mc}
    \end{center}
\end{figure}

In order to account for the stochastic nature of forming an OB association, our population synthesis model of an OB association is typically repeated 4000 times. This ensures to obtain a meaningful average for the temporal evolution of the nuclides abundance. As an example, the temporal evolution of the abundance of a few nuclides relevant to this work (\nuc{60}{Fe}, \nuc{56}{Fe} and \nuc{26}{Al}) is presented in Fig.~\ref{f:mc} for two OB associations having a total stellar mass of $10^4~M_\odot$ (red) and $10^5~M_\odot$ (blue). For clarity sake, five Monte Carlo realizations only are shown in black solid line for each case. The temporal evolution of the average mass of each nuclide computed for all realizations is represented as a solid colored line. As expected, the total mass of a given nuclide scales linearly with the stellar content of the OB association, and the variance of the nuclide mass distribution is larger for the OB association with the lowest stellar content (red case).

The temporal evolution of the mass of a nuclide in the gas of the OB association shows distinct behaviours depending on its lifetime. In case of stable nuclides (e.g. \nuc{56}{Fe}) the abundance increases monotonically with time as a result of the cumulative effect of successive nucleosynthetic events. For radioactive nuclides a typical saw tooth pattern is observed where sudden rises, corresponding to the enrichment of the OB association gas by the release of the wind and supernovae yields, are followed by the radioactive decay until another nucleosynthetic event builds up on top of the previous one. The obtained pattern depends on how the radionuclide lifetime compares with the mean time between two successive ccSN explosions $\Delta t_{ccSN}$ \citep{Cote2019}. In the case of the $10^4~M_\odot$ OB association $\Delta t_{ccSN}\approx0.5$~Myr. This is similar to the \nuc{26}{Al} lifetime ($\tau=1.03$~Myr) and the temporatal variation of its mass exhibits a larger scatter than for \nuc{60}{Fe} which has a longer lifetime ($\tau=3.78$~Myr). Since the stellar content of the $10^5~M_\odot$ OB association is higher, the mean time between two successive supernovae is lower ($\Delta t_{ccSN}\approx0.1$~Myr), and much smaller than both the \nuc{26}{Al} and \nuc{60}{Fe} lifetimes. In that case, the deviation between individual realizations (black curve) and the average (blue curve) is significantly reduced.

The nucleosynthetic enrichment of the gas of the OB association for a given nuclide may start at different epochs, as shown in Figure~\ref{f:mc}. When a nuclide is produced significantly by stellar winds (e.g. \nuc{26}{Al} and \nuc{56}{Fe}) it is enriching the OB association gas at early times. Since the wind contribution is released at explosion time in our model, the earliest possible release time occurs at $\sim3.5$~Myr which corresponds to the stellar lifetime of the most massive stars of our model ($120~M_\odot$). In the case of nuclides which are produced during the supernova phase only (e.g. \nuc{60}{Fe}) the first contributing stars are the exploding stars with shortest lifetime. This depends on the explodability criterion, which, in the present calculation, is such that stars with $M>25$~$M_\odot$ directly collapse to form black holes with no explosive contribution to the nucleosynthesis. The earliest release time in that case is $\sim7.7$~Myr which corresponds to the lifetime of a 25~M$_\odot$ star.

\subsection{CR production and transport}
Having assembled the interstellar content of \nuc{60}{Fe} nuclei within a group of stars, we proceed to determine the fraction ending up in locally-accelerated CRs, and propagate these then from the source through ISM toward the solar system.

\subsubsection{CR acceleration efficiency}
\label{sec:efficiency}

Galactic CRs are widely believed to be produced by the diffusive shock acceleration (DSA) process in SN remnants, but alternative sources such as massive star clusters, pulsar wind nebulae and the supermassive black hole at the Galactic center may also contribute to the CR population \citep[see][and references therein]{gabici2019}. The DSA theory predicts that a fraction of interstellar particles of about $10^{-5}$--$10^{-3}$ swept-up by a SN shock during the free expansion and the Sedov-Taylor phases become non-thermal, CR particles \citep[e.g.][]{blasi2013}. The CR-related gamma-ray luminosity of the Milky Way \citep{strong2010} suggests that the acceleration efficiency of protons, alpha-particles and other volatile elements is relatively low, of the order of $10^{-5}$ \citep{tatischeff2021}. But refractory elements such as Al and Fe are significantly more abundant than volatile ones in the CR composition compared to the solar system composition \citep{meyer1997}, which requires an acceleration efficiency of the former of the order of a few $10^{-4}$. 
Such higher efficiency could plausibly be explained by a more efficient injection of dust grains than ions into the DSA process, due to the higher rigidity of the former  \citep{ellison1997}. 

Massive star winds and SN ejecta within an OB association leave their sources in the form of hot, fast gas. As they expand, dust may form in dense clumps of stellar ejecta and condense a significant fraction of the refractory material. This has been suggested from infrared observations of SN~1987A \citep[e.g.][]{matsuura2019}. But some or all of this dust could be efficiently destroyed by thermal sputtering in the SN reverse shock. 
This is suggested from the paucity of presolar grains with characteristic signatures of core-collapse supernovae as analysed in meteoritic materials \citep{Nittler:1996,Hoppe:2019}. 
Subsequently, stellar ejecta are expected to be diluted in the hot superbubble plasma encompassing the stellar association. 
However, in a young and compact star cluster embedded in a molecular cloud, a fraction of the ejecta could be rapidly incorporated in cold molecular gas \citep{vasileiadis2013}. 
Gamma-ray observations of $^{26}$Al decay in nearby sources, such as the Scorpius-Centaurus and the Orion-Eridanus superbubbles, provide a unique way of studying the interstellar transport of massive star ejecta \citep[see][and references therein]{diehl2021}. 

The acceleration efficiency of massive star ejecta by SN shocks propagating into the superbubble plasma thus depends in theory on several parameters including the size and age of the parent OB association, as well as on the efficiencies of dust production in stellar ejecta and destruction by thermal sputtering. In our model, all these poorly-known processes are included in a single efficiency factor $\epsilon_{\rm acc}$, which we vary from $10^{-6}$ to $10^{-4}$. 

\subsubsection{CR propagation} \label{sec:propagation}
The general formalism of CR transport in the Galaxy includes particle diffusion, advection, ionization losses, spallation, and radioactive decay of unstable nuclei \citep{ginzburg1964}. The specific transport of $^{60}$Fe CRs has been recently studied by \cite{morlino2020} within the framework of a disk-halo diffusion model. They used for the CR diffusion coefficient, assumed to be the same in the disk and the halo \citep[see also][]{evoli2019}:  
\begin{equation}
D(R)= \beta D_0 \frac{(R/\text{GV})^{\delta}}{[1+ (R/Rb)^{\Delta \delta/s}]^{s}}~, 
\label{eq:diff}
\end{equation}
where $R$ is the particle rigidity, $D_0=3.08 \times 10^{28}$~cm$^{2}$~s$^{-1}$, $\delta=0.54$, $\Delta \delta=0.2$, $s=0.1$ and $R_b=312$ GV. For $^{60}$Fe CRs of $\approx 523$~MeV~nucleon$^{-1}$ (the mean LISM energy of the $^{60}$Fe nuclei detected by ACE/CRIS; see Sect.~\ref{sec:cris}), we have $D \approx 4.0 \times 10^{28}$~cm$^2$~s$^{-1}$. 

However, the diffusion coefficient in the local ISM is very uncertain. It depends in particular on the structure of the interstellar magnetic field between the nearby sources and the solar system. In addition, the spatial diffusion coefficient in an active superbubble environment is expected to be lower than that in the average ISM ($D_0$ in the range $10^{27}$--$10^{28}$~cm$^{2}$~s$^{-1}$; see \citet{vieu2022}). Moreover, in order to escape from a superbubble, CRs must diffuse mainly perpendicularly to the compressed magnetic field in the supershell, which could enhance the particle confinement in the hot plasma. Detailed modeling of these effects is beyond the scope of this paper. Here, we assume as a nominal value the same diffusion coefficient as \cite{evoli2019} and \cite{morlino2020}, and study in Sect.~\ref{sec:parameters} the impact on the results of reducing $D$ by an order of magnitude.

We now compare the timescales for the various processes involved in the transport of $^{60}$Fe ions in the Galactic disk, assuming the half-thickness of the disk to be $h = 150$~pc. With $D \approx 4.0 \times 10^{28}$~cm$^2$~s$^{-1}$ (as obtained from eq.~\ref{eq:diff}), the diffusion timescale of $^{60}$Fe CRs over this distance is
\begin{equation}
\tau_{\rm diff}= h^2 / D = 1.7 \times 10^5~{\rm yr}~, 
\end{equation}
which is significantly shorter than the CR advection timescale: 
\begin{equation}
\tau_{\rm adv}= h / u_0 = 2.9 \times 10^7~{\rm yr}~, 
\end{equation}
where $u_0 = 5$~kms~s$^{-1}$ is the typical CR advection velocity \citep{morlino2020}. Advection can thus be neglected.

The timescale for the catastrophic losses of $^{60}$Fe nuclei by nuclear spallation reactions in the ISM can be estimated as 
\begin{eqnarray}
\tau_{\rm spal} & = & \frac{1}{n_{\rm H} v [\sigma_{\rm H}+(n_{\rm He} /n_{\rm H})\sigma_{\rm He}]} \nonumber \\ 
& = & 1.54 \times 10^7 (n_{\rm H} / 0.1~{\rm cm^{-3}})^{-1}~{\rm yr}~, 
\label{eq:cata}
\end{eqnarray}
where $n_{\rm H}$ is the average ISM density into which $^{60}$Fe CRs propagate from their sources to the solar system, and $\sigma_{\rm H}$ and $\sigma_{\rm He}$ are the total reaction cross sections for fast ions propagating in interstellar H and He, respectively (we assume 90\% H and 10\% He by number). We used for these cross sections the universal parameterization of \citet{tripathi1996,tripathi1999}. We then found for the interaction mean free path of $^{60}$Fe nuclei of 523~MeV~nucleon$^{-1}$ in the ISM $\lambda_{\rm spal} = 2.69$~g~cm$^{-2}$, which is 5\% above the value reported by \citet{binns2016}: $\lambda_{\rm spal} = 2.56$~g~cm$^{-2}$. The total loss timescale of $^{60}$Fe CRs in the ISM is given by
\begin{equation}
\tau_{\rm loss}=\bigg(\frac{1}{\tau_{\rm spal}} + \frac{1}{\tau_{\rm decay}}\bigg)^{-1} ~, 
\label{e:tauLoss}
\end{equation}
where 
\begin{equation}
\tau_{\rm decay}=\gamma \tau_{\rm decay,0} = (5.90 \pm 0.09)\times 10^6~{\rm yr}~.
\end{equation}
Here $\gamma=1.56$ is the Lorentz factor and $\tau_{\rm decay,0}=3.78 \pm 0.06$~Myr is the mean lifetime for radioactive decay of $^{60}$Fe at rest. 

$^{60}$Fe ions originating from the younger subgroups of the nearby Sco-Cen OB association are expected to have propagated mainly in the low density gas ($n_{\rm H} \sim 0.1$~cm$^{-3}$) filling the Local Hot Bubble \citep{zucker2022}, and thus have suffered negligible catastrophic losses. But $^{60}$Fe ions coming from more distant OB associations (e.g. Orion, Cygnus OB2 etc..) and diffusing in the Galactic disk could have passed through denser regions (superbubble shells in particular) and seen on average ISM densities of $n_{\rm H} \sim 1$~cm$^{-3}$. However, $^{60}$Fe ions produced in distant associations should have mainly propagated in the low density halo of the Galaxy before reaching the solar system and then seen $n_{\rm H} \lsim 0.1$~cm$^{-3}$ \citep[see][]{morlino2020}. We thus adopt $n_{\rm H} = 0.1$~cm$^{-3}$ as the nominal value in our model, and will discuss the effect of changing the density parameter in Sect.~\ref{sec:parameters}. For $n_{\rm H} = 0.1$~cm$^{-3}$, $\tau_{\rm loss}=4.27$~Myr.

The ionization energy loss timescale for $^{60}$Fe ions of kinetic energy $E=523$~~MeV~nucleon$^{-1}$ is 
\begin{equation}
\tau_{\rm ion}= E / (dE/dt)_{\rm ion} = 5.7 \times 10^7 (n_{\rm H} / 0.1~{\rm cm^{-3}})^{-1}~{\rm yr}~, 
\label{eq:ion}
\end{equation}
where $(dE/dt)_{\rm ion}$ is the ionization energy loss rate, which is calculated from \citet[][eq. 4.24]{mannheim1994}. Like for the catastrophic energy losses, the significance of the ionization energy losses could depend on the OB association from which the $^{60}$Fe CRs originate. However, we see from Eqs.~\ref{eq:cata} and \ref{eq:ion} that $\tau_{\rm ion} > \tau_{\rm spal}$ whatever $n_{\rm H}$, so that the ionization losses can always be neglected in front of the catastrophic losses. 

So finally we consider a simple propagation model where accelerated ions, when escaping from their source, diffuse isotropically in the ISM and suffer both catastrophic and radioactive losses. We use as nominal set of input parameters $D_0=3.08 \times 10^{28}$~cm$^{2}$~s$^{-1}$ (Eq.~\ref{eq:diff}), $n_{\rm H} = 0.1$~cm$^{-3}$ and $\epsilon_{\rm acc}=10^{-5}$ (Sect.~\ref{sec:efficiency}), and we will study the impact of changing these parameters in Sec.~\ref{sec:parameters}. Future work could take into account in more detail the specific locations of the local OB associations and consider non-isotropic diffusion from MHD modeling of the LISM, but this is beyond the scope of the present paper. 

\subsubsection{CR density in the local ISM}
In order to compare the observed density of CRs by ACE/CRIS with our model we need to compute the number density of CRs $n(j)$ for a given nuclide $j$. This is computed as the sum of the contribution of each ccSN explosion from our model, where each SN accelerates with an efficiency $\epsilon_{\rm acc}$ (Sect.~\ref{sec:efficiency}) the number of atoms $N_0(j, t_i)$ of nuclide $j$ present at the explosion time $t_i$ of the $i^{\rm th}$ ccSN counted from the birth time of the OB association. This number of atoms is deduced from the temporal evolution of the mass of $j$ given in Eq.~\ref{e:mass} assuming that ccSNe do accelerate their own winds, since they are released prior to the collapse, but not their own ejecta \citep{Wiedenbeck1999}.  

CRIS measurements of $^{56}$Fe and $^{60}$Fe CRs were performed between $\sim 195$ and $\sim 500$~MeV~nucleon$^{-1}$, corresponding to $\Delta E \sim 400$--$700$~MeV~nucleon$^{-1}$ in the local ISM (Sect.~\ref{sec:cris}). From the Fe source spectrum obtained by \citet{boschini2021}, we find the fraction of Fe nuclei released with energies in the range $\Delta E$ to be $\epsilon_{\Delta E} \approx 5$\%. This quantity slightly depends on the assumed minimum CR energy used to calculate the total number of accelerated Fe. Thus, we have $\epsilon_{\Delta E} = 4.6$\% and $7.0$\% for $E_{\rm min}=1$ and $10$~MeV~nucleon$^{-1}$, respectively. 

The resulting CR population must then diffuse across the distance $d$ between the OB association and the solar system during a time $\Delta t_i = t_{\rm OB} - t_{i}$ where $t_{\rm OB}$ is the age of the association. The contribution of the $i^{\rm th}$ ccSN to the total number density is obtained from the solution of the diffusion equation and reads:
\begin{eqnarray}
    n_i(j) & = & \frac{N_0(j,t_i)\epsilon_{\rm acc} \epsilon_{\Delta E}(j)}{(4\pi D\Delta t_i)^{3/2}} \, \exp\left({-\frac{d^2}{4D\Delta t_i}}\right) \nonumber \\ 
     & \times & \exp\left(-\frac{\Delta t_i}{\tau_{\rm loss}(j)}\right),
    \label{e:diffusion}
\end{eqnarray}
where the last exponential decay term accounts for the catastrophic and radioactive losses (when $j$ is a radioactive species).  

\begin{figure}
    \centering
    \includegraphics[width=0.5\textwidth]{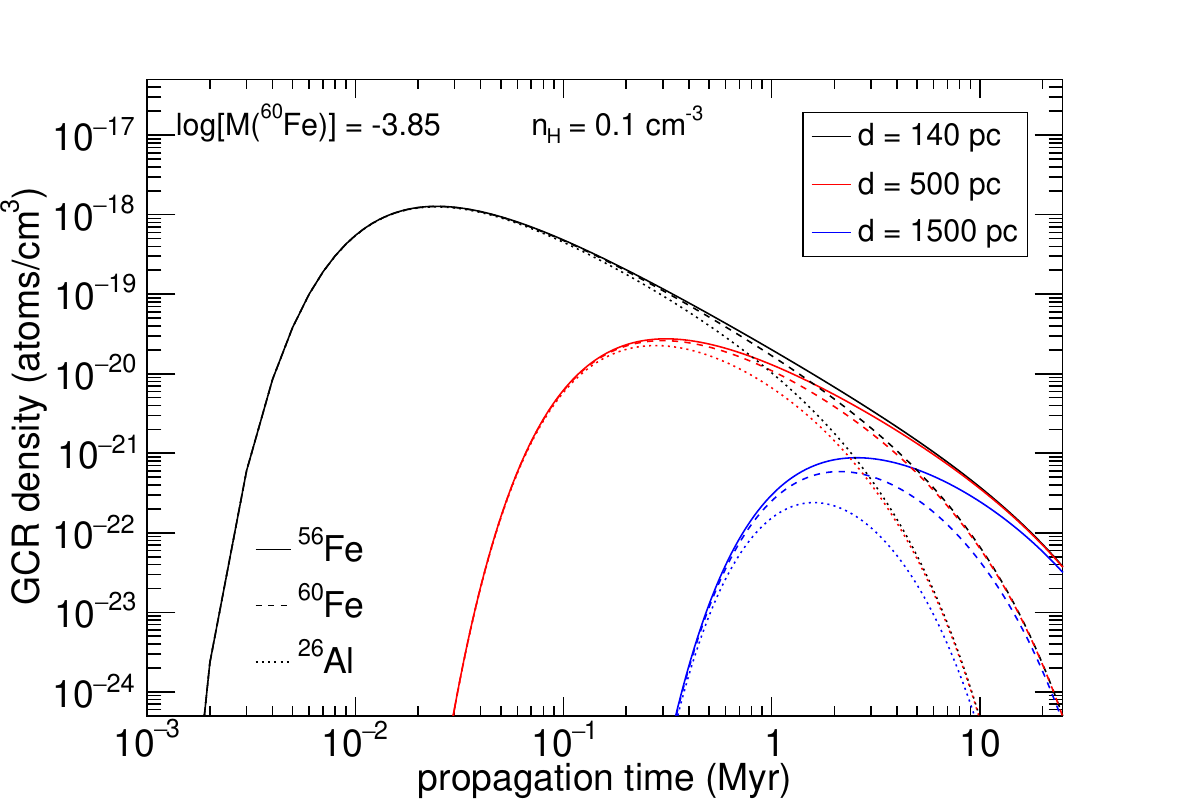}
    \caption{CR density for three different nuclides (\nuc{56}{Fe}, \nuc{60}{Fe} and \nuc{26}{Al}) as a function of propagation time for OB associations located at different distances $d$. The calculations are performed for a single ccSN event accelerating the same number of atoms for each nuclide corresponding to $1.4\times10^{-4}M_\odot$ of \nuc{60}{Fe}. All calculations are performed with a diffusion coefficient $D=4\times10^{28}$~cm$^2$ s$^{-1}$, a total efficiency $\epsilon_{\rm acc}\times\epsilon_{\Delta E} = 5\times10^{-7}$ and an average density of the ISM $n_{\rm H}=0.1$~cm$^{-3}$.}
    \label{f:diffusionGCR}
\end{figure}

The CR density obtained from Eq.~\ref{e:diffusion} is displayed in Fig.~\ref{f:diffusionGCR} as a function of the propagation time $\Delta t_i$ for three different distances $d$ of the parent OB association. Calculations are performed for one stable nuclide (\nuc{56}{Fe}) and two radionuclides (\nuc{60}{Fe} and \nuc{26}{Al}). For all cases we consider for illustration purpose the same number of atoms in the parent superbubble plasma, $N_0=2.8\times 10^{51}$, which corresponds to $1.4\times10^{-4} M_\odot$ of \nuc{60}{Fe}. This value is obtained from an average of the \cite{LC18} yields over the IMF from \cite{Kroupa2001} and the initial rotational velocity from \cite{Prantzos2018}. Fig.~\ref{f:diffusionGCR} exhibits the expected time evolution of the CR number density at the solar system location from sources at various distances, with a sharp rise and a longer decay. For the low average ISM density considered, $n_{\rm H}=0.1$~cm$^{-3}$, the catastrophic losses are negligible wrt the radioactive decay losses of \nuc{26}{Al} and \nuc{60}{Fe}, which explains why for $\Delta t_i \gsim 1$~Myr the density of \nuc{26}{Al} ($\tau_{\rm decay,0}=1.03$~Myr) decreases faster than that of \nuc{60}{Fe} ($\tau_{\rm decay,0}=3.78$~Myr) and \nuc{56}{Fe}. 

We see in Fig.~\ref{f:diffusionGCR} that the CR density at maximum varies a lot with the source distance, e.g. by more than three orders of magnitude from $d=140$~pc (the  approximate distance of the Sco-Cen association) to $d=1.5$~kpc (the approximate distance of Cyg OB2). The time when the CR density reaches its maximum can be obtained after canceling the derivative of $n_i(j)$ from Eq.~\ref{e:diffusion}: 
\begin{equation}
    t_{\rm max} = \frac{\sqrt{D\tau(4d^2 + 9D\tau)} - 3D\tau}{4D}~.
    \label{e:tmax}
\end{equation}
For stable nuclei and when catastrophic losses are negligible, we retrieve the well-known formula:
\begin{equation}
    t_{\rm max} = \frac{d^2}{6D}.
    \label{e:tmaxStable}
\end{equation}
For $d=140$~pc and $D=4\times10^{28}$~cm$^2$ s$^{-1}$, we get $t_{\rm max}=25$~kyr, which is much shorter than $\tau_{\rm loss}$ for both \nuc{26}{Al}, \nuc{60}{Fe} and \nuc{56}{Fe}. But for $d=1.5$~kpc, $t_{\rm max}$ is comparable to the radioactive lifetime of \nuc{26}{Al} and \nuc{60}{Fe}, which thus have time to decay before reaching the solar system.

\section{Nearby OB associations} \label{sec:OB}
The nearest OB associations have been identified and studied since a long time \citep[e.g.][]{blaauw1964}. The catalogue from \cite{deZeeuw1999} based on $Hipparcos$ positions, proper motions and parallaxes provides a census of the stellar content of the OB associations within 1~kpc from the Sun. With improved astrometry, $Gaia$ allows a better determination of the membership of stars belonging to OB associations, and the identification of new sub-groups \citep{Zucker2022a}. Recent compilations of O and B stars \citep{pantaleoni2021} and OB associations \citep{Wright2020} make use of $Gaia$'s results.

In the present work we consider all the \textit{well-studied OB associations} listed in \cite{Wright2020} and all the \textit{high-confidence OB associations} at less than 1~kpc. Properties of these OB associations are summarized in Table~\ref{t:ob} where their distance and age come from the review of \cite{Wright2020}. The numbers of observed stars mainly come from the catalogs of \cite{deZeeuw1999} and \cite{melnik2017} except for a few OB associations for which the star census has been extensively studied such as Orion OB1 \citep{brown1994,hillenbrand1997}, Perseus OB2 \citep{Belikov2002} and Vela OB2 \citep{armstrong2018}.

\begin{table*}
    \caption{Properties of local OB associations considered in the present work. Distances and ages come from \protect\cite{Wright2020}. The richness ($N_* \geq 8 M_{\odot}$ at formation time) and the number of core collapse supernovae having already exploded ($N_{ccSN}$) are calculated from the number of observed stars and the OB association age (see text). $N_{ccSN}$ is reported for two different explodability criterion.}
    \centering
    \begin{tabular}{lccccccc}
        \hline \hline
        Association & Distance (pc) & Age (Myr) & Number of observed stars & Ref. & $N_*(\geq 8 M_{\odot})$ & $N_{ccSN}^{LC18}$ & $N_{ccSN}^{Sukhbold}$ \\ \hline
        Sco-Cen: US   & $143\pm6$   & $10\pm7$   & $N(OB)=49$              & DZ99 & $13.9\pm3.5$ & $1.8\pm1.8$  & $1.6\pm1.4$ \\
        Sco-Cen: UCL  & $136\pm5$   & $16\pm7$   & $N(OB)=66$              & DZ99 & $19.4\pm4.3$ & $5.1\pm3.2$  & $4.2\pm2.7$ \\
        Sco-Cen: LCC  & $115\pm4$   & $15\pm6$   & $N(OB)=42$              & DZ99 & $12.4\pm3.3$ & $3.1\pm2.1$  & $2.5\pm1.7$ \\ \hline
        Ori OB1a      & $\sim360$   & $8-12$     & $N(4-15M_\odot)=53$     & B94  & $26.5\pm4.9$ & $2.2\pm1.6$  & $2.2\pm1.4$ \\
        Ori OB1b      & $360-420$   & $2-8$      & $N(4-120M_\odot)=45$    & B94  & $19.4\pm3.6$ & $0.5\pm0.1$  & $0.8\pm0.6$ \\
        Ori OB1c      & $\sim385$   & $2-6$      & $N(7-36M_\odot)=23$     & B94  & $22.7\pm2.5$ & 0            & $0.8\pm0.6$ \\ 
        Ori OB1d      & $\sim380$   & $1-2$      & $N(>1M_\odot)=145$      & H97  & $10.0\pm3.0$ & 0            & 0           \\ \hline
        Per OB1       & $\sim1830$  & $8-11$     & $N(OB)=133$             & MD17 & $36.1\pm5.6$ & $2.2\pm1.6$  & $2.6\pm1.1$ \\ 
        Per OB2       & $296\pm17$  & $1-10$     & $N(1-17M_\odot)=800$    & B02  & $53.9\pm7.2$ & $0.9\pm1.1$  & $1.8\pm1.5$ \\
        Per OB3       & $175\pm3$   & $50$       & $N(OB)=30$              & DZ99 & $10.8\pm3.7$ & $7.9\pm3.4$  & $7.1\pm3.2$ \\ \hline
        Cyg OB2       & $1350-1750$ & $1-7$      & $N(O)=78$               & B20  & $241.2\pm47.8$ & 0  & $3.8\pm3.4$ \\ 
        Cyg OB4       & $\sim800$   & $\sim 8.3$ & $N(OB)=2$               & MD17 & $1.3\pm0.8$  & $0.5\pm0.2$  & $0.5\pm0.2$ \\
        Cyg OB7       & $\sim630$   & $1-13$     & $N(OB)=25$              & MD17 & $7.2\pm2.4$  & $0.7\pm0.6$  & $0.8\pm0.6$ \\
        Cyg OB9       & $\sim960$   & $2-4$      & $N(OB)=31$              & MD17 & $8.5\pm2.5$  & 0            & $0.5\pm0.2$ \\ \hline
        Vel OB2       & $411\pm12$  & $10-30$    & $N(>2.5M_\odot)=72$     & A18  & $18.4\pm4.4$ & $6.8\pm3.9$  & $5.6\pm3.5$ \\ \hline
        Trumpler 10   & $372\pm23$  & $45-50$    & $N(OB)=22$              & DZ99 & $8.3\pm3.2$  & $6.7\pm2.8$  & $6.1\pm2.6$  \\ \hline
        Cas-Tau       & $125-300$   & $\sim50$   & $N(OB)=83$              & DZ99 & $29.4\pm6.6$ & $22.1\pm6.5$ & $20.0\pm6.2$ \\ \hline
        Lac OB1       & $368\pm17$  & $2-25$     & $N(OB)=36$              & DZ99 & $10.7\pm3.1$ & $2.5\pm2.3$  & $2.2\pm2.0$ \\ \hline
        Cep OB2       & $\sim730$   & $5$        & $N(OB)=56$              & DZ99 & $15.5\pm3.4$ & $0$          & $0.8\pm0.5$ \\
        Cep OB3       & $\sim700$   & $5-8$      & $N(OB)=25$              & MD17 & $7.2\pm2.4$  & $0.5\pm0.1$  & $0.7\pm0.4$ \\
        Cep OB4       & $\sim660$   & $1-6$      & $N(OB)=7$               & MD17 & $2.5\pm1.2$  & $0$          & $0.5\pm0.2$ \\
        Cep OB6       & $270\pm12$  & $\sim50$   & $N(OB)=6$               & DZ99 & $2.9\pm1.8$  & $2.3\pm1.5$  & $2.1\pm1.4$ \\\hline
        Collinder 121 & $543\pm23$  & $5$        & $N(OB)=87$              & DZ99 & $22.9\pm4.1$ & $0$          & $0.9\pm8.8$ \\ \hline
        Cam OB1       & $\sim 800$  & $7-14$     & $N(OB)=45$              & MD17 & $12.9\pm3.3$ & $1.6\pm1.3$  & $1.5\pm1.1$ \\ \hline
        Mon OB1       & $\sim 580$  & $1-10$     & $N(OB)=6$               & MD17 & $2.3\pm1.2$  & $0.5\pm0.1$  & $0.5\pm0.2$ \\
        \hline \hline
    \end{tabular}
    \\
   {References: A18 \citep{armstrong2018}, B02 \citep{Belikov2002}, B20 \citep{Berlanas2020}, B94 \citep{brown1994}, DZ99 \citep{deZeeuw1999}, H97 \citep{hillenbrand1997}, MD17 \citep{melnik2017}.}
    \label{t:ob}
\end{table*}

We compute in the 6$^{\rm th}$ column of Table~\ref{t:ob} the richness of the OB association (or subgroup) which we define as the number of massive stars ($\geq8M_\odot$) present when the association is formed. It is estimated considering the number of observed stars and the age of the OB association. For each realization of our population synthesis model an OB association is first given an age obtained by uniformly sampling the range of adopted ages (Table~\ref{t:ob}, 3$^{rd}$ column). We assume 50\% uncertainty on the age when this is not specified. In a second step, the IMF \cite[from][]{Kroupa2001} is sampled until the number of observed stars is reproduced taking into account the star lifetime from \cite{LC18}. The number of massive stars is then recorded for each realization; the richness and associated standard deviation are obtained for typically 4000 realizations.

The determination of the richness depends on the mass range associated to the number of observed stars. However, this is only reported for very few OB associations: Orion OB1, Perseus OB1 and OB3, and Vela OB2. When the number of O and B stars is given instead, we use for the latest B-type stars a mass of $2.8~M_\odot$ obtained from a study of binary systems \citep{habets1981}. A similar value is obtained using the evolutionary tracks from \cite{palla1999} for pre-main-sequence models as shown in \cite{Preibisch2002}. In these conditions, the total number of stars we obtain for Upper Scorpius when normalizing the IMF from \cite{Preibisch2002} to the 49 B stars reported in \cite{deZeeuw1999} is 2590, in very good agreement with the 2525 stars reported by \cite{Preibisch2002}. For the latest O-type stars we consider that they have masses of $16~M_\odot$ or larger \citep{habets1981,Weidner2010}. 

The past nucleosynthetic activity of an OB associations is related to the number of ccSN that have exploded so far ($N_{ccSN}$). In the same calculation as for the richness, the number of exploding massive stars with a stellar lifetime shorter than the age of the OB association is recorded. We use by default the explodability criterion from \cite{LC18}, i.e. $M<25M_\odot$, and the corresponding $N_{ccSN}$ is reported in the 7$^{\rm th}$ column of Table~\ref{t:ob}. For young OB associations with ages smaller than the first ccSN explosion time (occurring at about 7.8~Myr, and corresponding to the lifetime of a non rotating 25~$M_\odot$ star) no ccSN have exploded yet. In these cases the enrichment of the gas of the OB association mainly comes from stellar winds which cannot be accelerated as CRs because no supernova exploded yet. Hence, the OB association is not expected to contribute to the CR density budget even though it may be a high richness association (e.g. Cyg OB2, Orion OB1c, Collinder 121). On the contrary, for rather old OB associations with ages greater than the last massive star explosion time (occurring at about 40~Myr, and corresponding to the lifetime of a non rotating 8~$M_\odot$ star) all massive stars may have exploded. However, even in the case of a high richness OB association (e.g. Cas-Tau), the present day enrichment in short lived radionuclides (e.g. \nuc{26}{Al}, \nuc{60}{Fe}) of the associated superbubble gas will most likely be negligible because of the smaller yields for the low-end massive stars and the free radionuclide decay after the last massive star explosion (see Fig.~\ref{f:mc}).

The number of past ccSN for a given OB association significantly depends on the explodability criterion which is considered. In the last column of Table~\ref{t:ob} we compute $N_{ccSN}$ using \cite{Sukhbold2016} explodability criterion. In this case, and at variance with the case using \cite{LC18} explodability criterion, some stars having initial masses greater than 25~$M_\odot$ explode as ccSN. These stars have lifetime smaller than 7.8~Myr, so younger OB associations will have a nucleosynthetic activity while it is not the case with \cite{LC18} explodability criterion (e.g. Orion OB1c). On the other hand, for older OB associations the nucleosynthetic activity may be reduced when considering the \cite{Sukhbold2016} explosion criterion (e.g Sco-Cen) since some stars in the low-end of the massive range ($<25~M_\odot$) may not explode as ccSN.

The number of past ccSN we obtain for Sco-Cen is $10\pm7.1$ or $8.3\pm5.8$ depending on the explodability criterion, which is in reasonable agreement with the number of past supernovae, between 14 and 20, needed to excavate the Local Bubble \citep{Fuchs2006,Breitschwerdt2016}.

\begin{figure*}
    \centering
    \includegraphics[width=0.99\textwidth,
    clip]{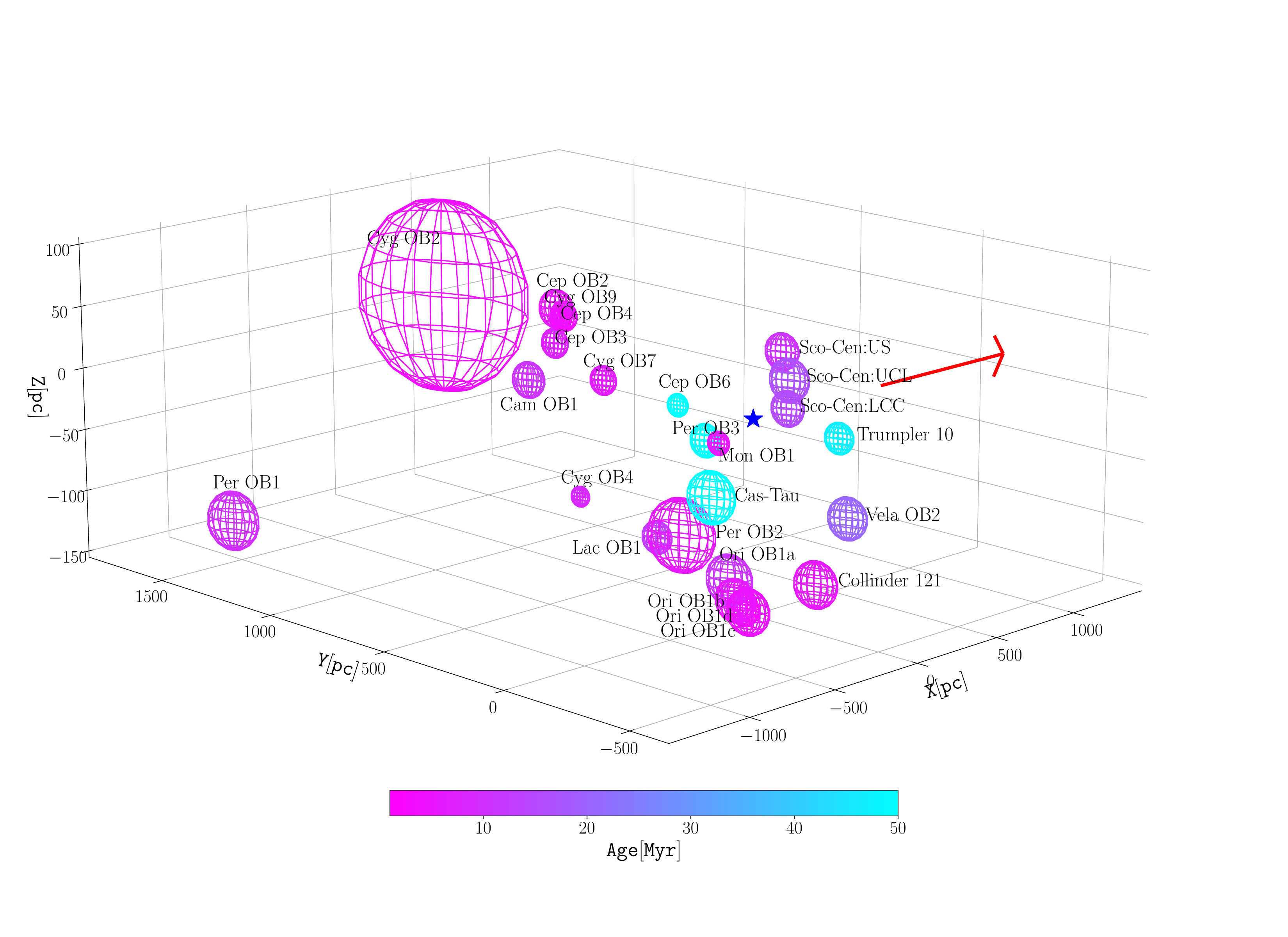}
    \caption{3D representation of the local OB associations in the solar neighborhood. The volume of each sphere is proportional to the richness of the association, the color accounts for the typical age of the association~(see Table~\ref{t:ob}). The blue star indicates the position of the Sun and the red arrow points towards the Galactic center.}
    \label{f:OB}
\end{figure*}

A 3D representation of the OB associations in our solar neighbourhood is presented in Fig.~\ref{f:OB} with the volume of each OB association proportional to its richness while the age of the association is color coded.

\section{Origin of $^{56,60}$Fe and $^{26}$Al in CRs} \label{sec:results}
\subsection{CR density distribution and observations}
\begin{figure*}
    \includegraphics[width=1.02\textwidth]{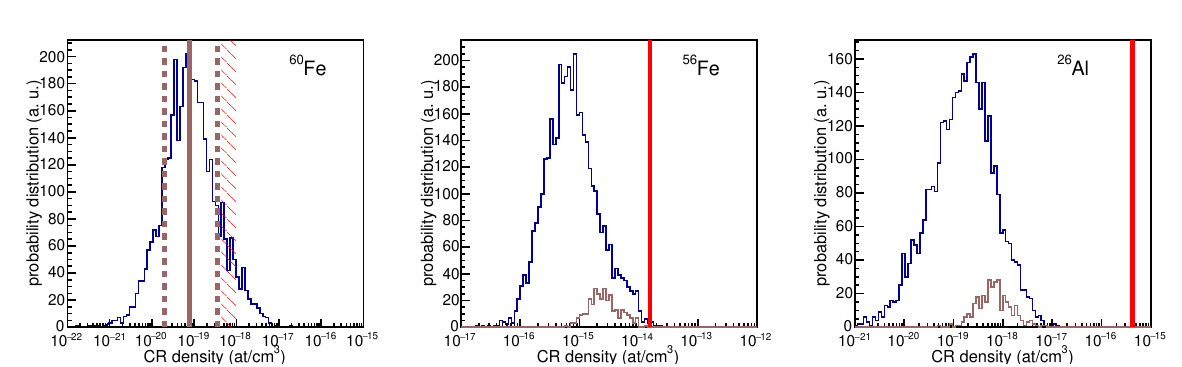}
    \caption{Calculated CR density distribution in the LISM (blue histograms) for $N=4000$ realizations of our model including all OB associations listed in Table~\ref{t:ob} for \nuc{60}{Fe} (left panel), \nuc{56}{Fe} (middle panel) and \nuc{26}{Al} (right panel). Vertical brown thick lines correspond to the 16$^{\rm th}$, 50$^{\rm th}$ and 84$^{\rm th}$ percentile of the \nuc{60}{Fe} CR density distribution. Hatched and solid vertical red lines correspond to the CR densities derived from the ACE/CRIS measurements. Brown histograms represent cases for which the predicted \nuc{60}{Fe} CR density matches the observations. Nuclides in the gas of the OB associations are injected in CRs with an efficiency $\epsilon_{acc}=10^{-5}$. CRs are assumed to propagate in the ISM of average density $n_{\rm H}=0.1$~cm$^{-3}$ with a mean diffusion coefficient $D_0=3.08\times10^{28}$~cm$^2$ s$^{-1}$ (eq.~\ref{eq:diff}).}
    \label{f:gcrDensity}
\end{figure*}

Our CR population synthesis model was used to compute the LISM CR density of \nuc{60}{Fe}, \nuc{56}{Fe} and \nuc{26}{Al} resulting from the contribution of all OB associations listed in Table~\ref{t:ob}. We take as nominal parameters for the CRs acceleration and propagation a mean acceleration efficiency $\epsilon_{acc}=10^{-5}$, a mean diffusion coefficient $D_0=3.08\times10^{28}$~cm$^2$ s$^{-1}$ (eq.~\ref{eq:diff}) and a mean ISM average density $n_{\rm H}=0.1$~cm$^{-3}$. The impact of these parameters will be discussed in Sec.~\ref{sec:parameters}. A realization of our CR population synthesis model is defined as the sampling of the IMF until the richness of each considered OB associations is reproduced. For each realization, the distance of an OB association needed to calculate the CR densities is obtained by uniformly sampling the adopted distances (Table~\ref{t:ob}, $2^{\rm nd}$ column). We assume 15\% error on the distance when the uncertainty is not specified. The total CR density distribution for a given nuclide, obtained as the sum of the contribution of each OB association, is shown in Fig.~\ref{f:gcrDensity} (blue histogram) for 4000 realizations, and compared to the ACE/CRIS measurements (hatched and solid vertical red lines). The median of the total \nuc{60}{Fe} CR density distribution is indicated with the solid brown vertical lines and the 16$^{\rm th}$ and 84$^{\rm th}$ percentiles, defining a 68\% probability coverage, are the dashed brown vertical lines.

The calculated \nuc{60}{Fe} CR density has a rather broad distribution with a mean of $\approx2.5\times10^{-19}$~atoms cm$^{-3}$. This is about 2.8 times smaller than what is deduced from the ACE/CRIS observations \citep{binns2016}. However, the observations are well within the calculated distribution at slightly more than $1\sigma$ from the median. This indicates that the observed density of \nuc{60}{Fe} CRs in the LISM is not exceptional, indeed it represents $\approx8\%$ of the simulated cases. The spread of the distribution arises from the stochastic nature of the IMF, the different \nuc{60}{Fe} yields as a function of the stellar initial mass, the competition between the \nuc{60}{Fe} lifetime and the mean time between two successive ccSN, and the contribution from the different OB associations.

The calculated \nuc{56}{Fe} CR density distribution (blue) is not as broad as in the case of \nuc{60}{Fe} which is due to the similar \nuc{56}{Fe} yield for each ccSN and the stable nature of \nuc{56}{Fe}. The CR distribution for the realizations matching the \nuc{60}{Fe} observations is also displayed as a brown histogram. On average, the calculated \nuc{56}{Fe} CR density represents about 20\% of the observed value. This suggests that a non negligible fraction of the \nuc{56}{Fe} CR density in the LISM comes from local sources (see further discussion in Sec.~\ref{sec:locality}).

Concerning the calculated CR density for \nuc{26}{Al} it is lower by more than one order of magnitude than the ACE/CRIS observations. This is expected since \nuc{26}{Al} is mostly produced by CRs spallation \citep{yanasak2001} and that our CR population synthesis model only computes the primary component of CRs. This result suggests that the excess in the Al CR spectrum found by \cite{boschini2022} is not produced by a contribution of primary \nuc{26}{Al}.

\subsection{The role of Sco-Cen}
\begin{figure}
    \includegraphics[width=0.5\textwidth]{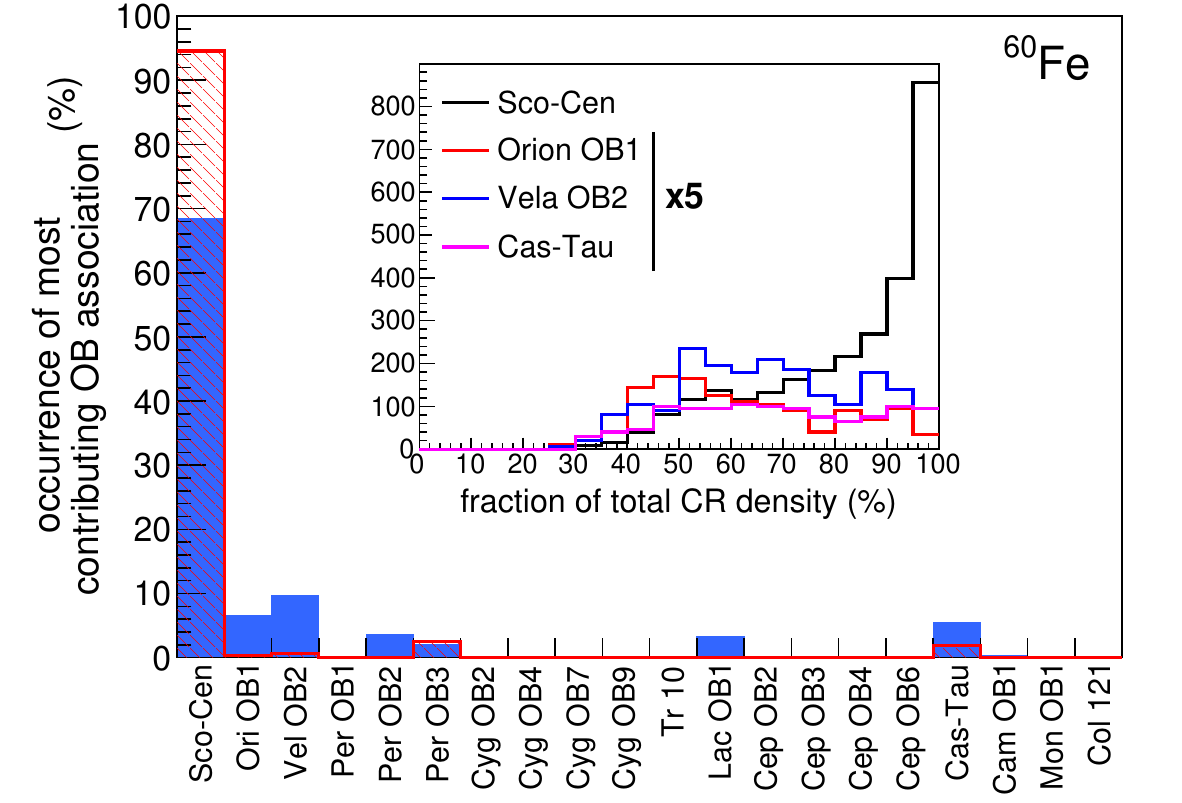}
    \caption{Occurrence of the most contributing OB association to the total CR density of \nuc{60}{Fe} in the LISM (solid blue) for $N=4000$ realizations of our CR population synthesis model. The hatched red histogram is the same but for realizations matching the \nuc{60}{Fe} ACE/CRIS observations. The distribution of the fraction of total CR density for the main contributing OB associations is given in the inset; the distribution for OB associations other than Sco-Cen is displayed with a scaling factor of 5.}
    \label{f:gcrRelativeContribution}
\end{figure}

For each realization of our CR population synthesis model it is interesting to know which OB association is contributing the most to the total CR density of \nuc{60}{Fe} (shown in Fig.~\ref{f:gcrDensity}). This is what is represented by the blue histogram in Fig.~\ref{f:gcrRelativeContribution} which indicates that in $\approx70\%$ of the cases Sco-Cen is the main contributor to the predicted total CR density of \nuc{60}{Fe}, followed by Vela OB2, Orion OB1 and Cas-Tau at the 10\% level. However, this does not tell anything about whether, realization by realization, the most contributing OB association dominates largely the other ones or whether its contribution is more equally shared. The inset in Fig.~\ref{f:gcrRelativeContribution} shows the distribution of the fraction of total CR density for the most contributing OB associations. The distributions are very different between Sco-Cen and the other associations. For Sco-Cen the distribution is peaked for large fractions meaning that when Sco-Cen is the main contributing association this is by far the dominant one. Specifically, Sco-Cen contributes by more than 80\% to the total CR density in 64\% of the cases. For Vela OB2, Orion OB1 and Cas-Tau the fraction distributions are rather flat with a maximum at about $50\%-60\%$ indicating that the contribution to the total CR density is much more equally shared between the participating associations.

\begin{figure}
    \includegraphics[width=0.5\textwidth]{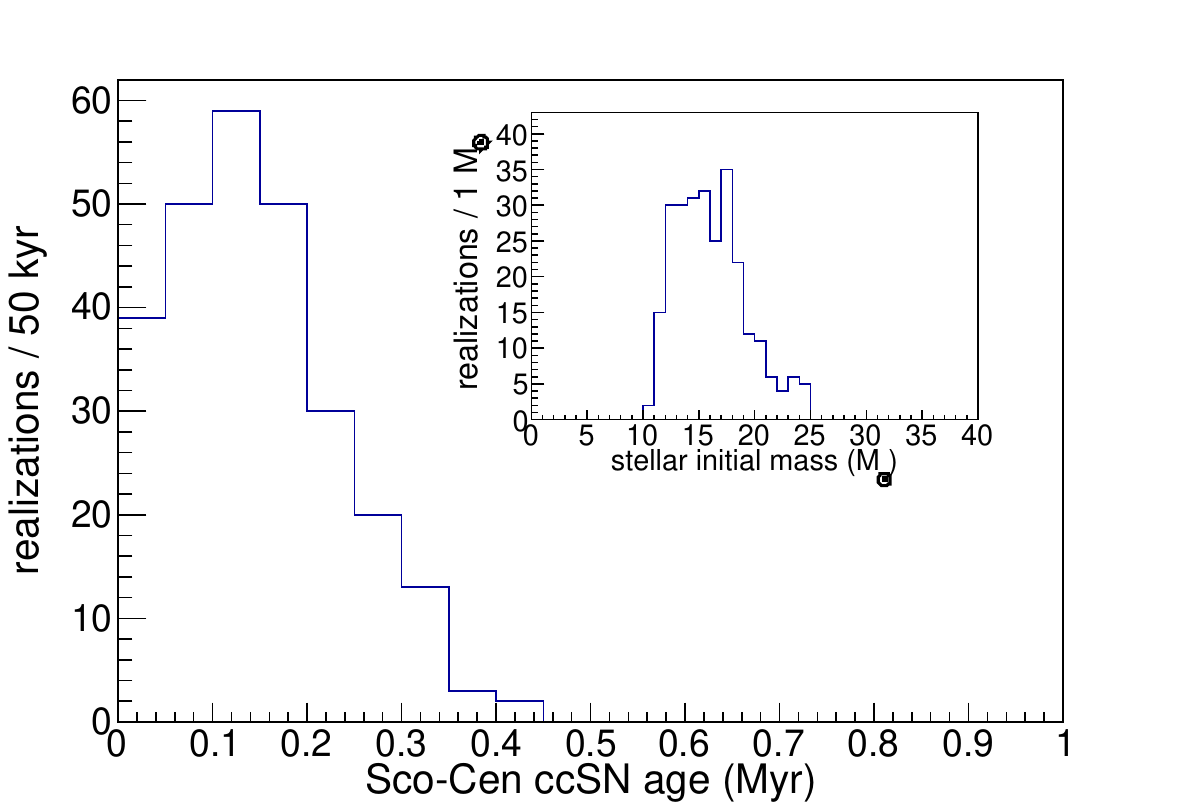}
    \caption{Age and stellar initial mass distributions for the Sco-Cen supernova accelerating more than 50\% of the observed \nuc{60}{Fe} CR density measured by ACE/CRIS.}
    \label{f:scoCenConfig}
\end{figure}

If we now only consider the realizations compatible with the ACE/CRIS observations, it appears that the 15 detected \nuc{60}{Fe} nuclei are nearly always coming from the Sco-Cen association as shown by the red hatched histogram in Fig.~\ref{f:gcrRelativeContribution}. The configuration of these realizations is quite specific since they all involve at least one massive star having exploded recently. Fig.~\ref{f:scoCenConfig} shows the explosion time distribution for the realizations where a single supernova in Sco-Cen accelerates more than 50\% of the total \nuc{60}{Fe} CR density. The mean explosion time is 146~kyr with a RMS of 90~kyr, and in 93\% of the cases the age of the supernova is smaller than 300~kyr. The initial mass distribution of these supernovae at the origin of the acceleration of the \nuc{60}{Fe} present in the enriched gas of the OB association is represented in the inset of Fig.~\ref{f:scoCenConfig}. It is characterized by a mean stellar initial mass of $16.1~M_\odot$ with a RMS of $3.2~M_\odot$.

\section{Discussion} \label{sec:discussion}
\subsection{Sensitivity to Input Parameters} \label{sec:parameters}
The results presented so far are obtained with a nominal set of input parameters. However, the underlying physics of both the population synthesis and the acceleration and propagation of CRs is far from being under control. It is thus important to explore the robustness of our findings by varying the main ingredients of our CR population synthesis model within uncertainties. Concerning the population synthesis part of our model, we first investigate a case where the IMF is determined from the observed Upper Scorpius OB association members \citep{Preibisch2002} rather than from the model developed by \citet{Kroupa2001}. We also investigate a case with the explodability criterion taken from \cite{Sukhbold2016} and not from \citet{LC18}.  Finally we perform a simulation with the yields obtained from the PUSH model \citep{Curtis2019} based on the pre-explosion models of \cite{WH07} for non rotating solar metallicity stars. 

\begin{figure}
    \includegraphics[width=0.5\textwidth]{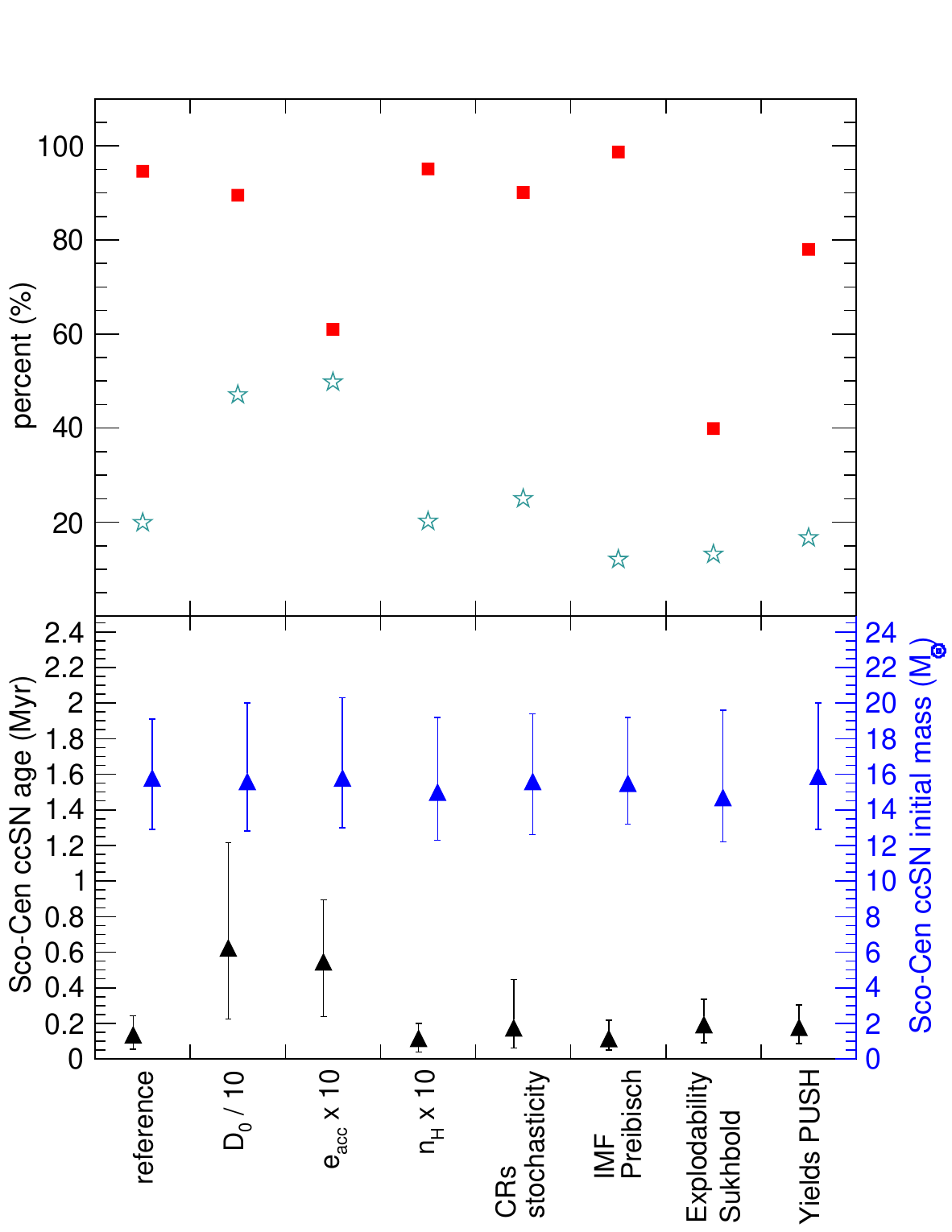}
    \caption{Effect of the input parameters of our CR population synthesis model when we consider the realizations matching the \nuc{60}{Fe} observations from ACE/CRIS. {\it Top:} occurrence of Sco-Cen as the most contributing OB association (filled red squares), and ratio of the calculated mean \nuc{56}{Fe} CR density to the observed one (open star symbols). {\it Bottom:} age (black) and mass (blue) of the Sco-Cen supernova accelerating more than 50\% of the observed \nuc{60}{Fe} CR density.}
    \label{f:sensitivity}
\end{figure}

Concerning the acceleration and propagation of CRs, we explore four different cases by varying the relevant input parameters. First, we consider a case where the diffusion coefficient is reduced by an order of magnitude, i.e. taking $D_0=3.08 \times 10^{27}$~cm$^{2}$~s$^{-1}$ in Eq.~\ref{eq:diff}, which may be expected if CRs spent most of their time in an active superbubble environment (Sec.~\ref{sec:propagation}). We also consider a case where the average density of the ambient medium is increased by an order of magnitude, i.e. $n_{\rm H}=1~$cm$^{-3}$, which may be typical if CRs diffuse for a significant time in superbubble shells and/or in the ISM of the Galactic disk outside superbubbles. We also study the impact of increasing the CR acceleration efficiency by an order of magnitude, i.e. $\epsilon_{\rm acc}=10^{-4}$, to take into account that refractory elements such as Fe may be more efficiently accelerated by the DSA process than volatile elements (Sec.~\ref{sec:efficiency}). Finaly, we compute a case where $D_0$, $n_{\rm H}$ and $\epsilon_{\rm acc}$ are independently determined for each model realization from a log-normal distribution with a factor uncertainty of 2. This case is intended to take into account that the acceleration of CRs and their propagation to the solar system could depend on the individual properties of the nearby OB associations and their specific location in the ISM. In particular, $\epsilon_{\rm acc}$ could depend on the size and age of the parent OB association, whereas $D_0$ and $n_{\rm H}$ could be related to the distance of the source and its position wrt the magnetic field lines passing near the solar system. Results are gathered in Fig.~\ref{f:sensitivity} for the realizations matching the LISM \nuc{60}{Fe} CR density determined from the ACE/CRIS observations. 

\begin{figure}
    \includegraphics[width=0.5\textwidth]{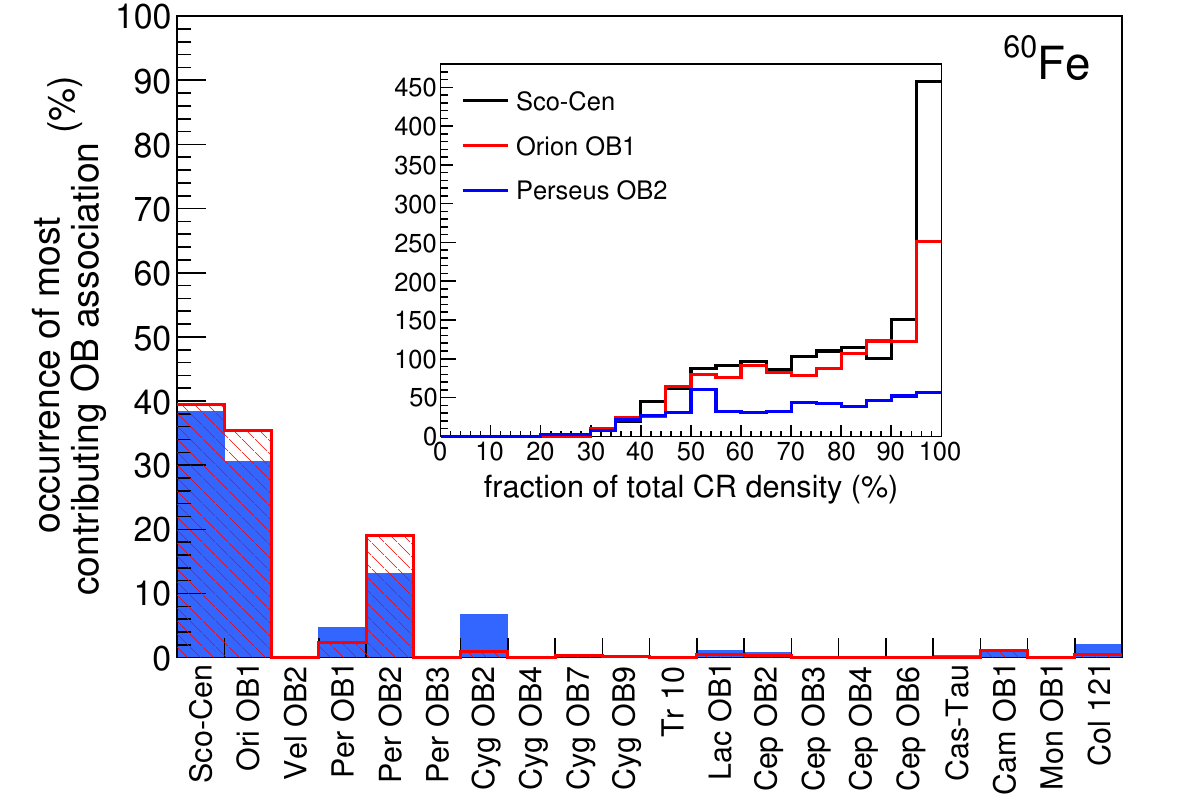}
    \caption{Same as Fig.~\ref{f:gcrRelativeContribution} except that the explodability criterion is from \protect\cite{Sukhbold2016}.}
    \label{f:gcrRelativeContributionSukhbold}
\end{figure}

The red full squares represent the probability that the observed LISM \nuc{60}{Fe} CR density can be explained by the contribution of the Sco-Cen OB association only. We see that the results are largely independent of the assumed input parameters except for the acceleration efficiency and the explodability criterion. In the later case the Orion OB1 and Perseus OB2 associations are also able, on their own, to produce the observed \nuc{60}{Fe} CR density (see Fig.~\ref{f:gcrRelativeContributionSukhbold}). This can be semi-quantitatively explained by comparing the number of past supernovae ($N_{ccSN}$) reported in the last two columns of Table~\ref{t:ob}. Indeed, the ratio $N_{ccSN}$(Sco-Cen)/$N_{ccSN}$(Orion OB1) decreases from $10.0/2.7=3.7$ to $8.3/3.8=2.2$ when considering the \cite{LC18} or \cite{Sukhbold2016} explodability criterion, respectively. This indicates that the relative contribution of Orion OB1 should increase with respect to Sco-Cen when the \cite{Sukhbold2016} explosion criterion is considered, as observed in Fig.~\ref{f:gcrRelativeContributionSukhbold}. In a similar way, the number of past supernovae is higher (lower) for Perseus OB2 (Vela OB2) when considering the \cite{Sukhbold2016} explodability criterion, which consequently increases (reduces) the importance of these OB associations with respect to the reference case using the \cite{LC18} explodability criterion.

In the case where the acceleration efficiency is multiplied by a factor of 10, the Sco-Cen association still dominates over the other associations, but with a reduced probability of about 60\%. Indeed, with a higher acceleration efficiency more material is accelerated resulting in an increased predicted \nuc{60}{Fe} CR density. It follows that OB associations other than Sco-Cen are then also able to account for the observed \nuc{60}{Fe} CR density. There are then a larger number of realizations where a smaller contribution from Sco-Cen due to the initial draw of the massive star population is compensated by a larger contribution from associations such as Orion OB1 or Perseus OB2. This consequently reduces the occurrence of Sco-Cen as the most contributing OB association.

The green open stars in Fig.~\ref{f:sensitivity} represent the ratio between the average value of the calculated \nuc{56}{Fe} CR density for the realizations which account for the \nuc{60}{Fe} observations, and the observed \nuc{56}{Fe} CR density value. This can be taken as an indicator of the \nuc{56}{Fe} CR density fraction which may come from a local source. This fraction is typically 20\% except for the cases where the CR diffusion coefficient and acceleration efficiency are varied by a factor of 10; for such cases the fraction is about 50\%. Since the nucleosynthesis activity for these two cases is the same, the higher predicted \nuc{56}{Fe} CR density is correlated to an increase of the \nuc{60}{Fe} CR density. The reason for such an increase when the CR acceleration efficiency is higher has previously been discussed. When the CR diffusion coefficient is decreased, the propagation time needed to reach the maximum CR density for a given OB association distance is higher (see Eq.~\ref{e:tmaxStable}). Thus, there are possibly more ccSN which can explode during this time lapse, leading then to a higher predicted \nuc{56,60}{Fe} CR density.

Since in most cases Sco-Cen is the OB association contributing the most to the LISM \nuc{60}{Fe} CR density, it is interesting to explore the impact of the input parameters on the properties of the supernovae accelerating the material present in the gas of the superbubble. Fig.~\ref{f:sensitivity} (bottom) shows the age (black) and mass (blue) of the Sco-Cen ccSN accelerating more than 50\% of the observed \nuc{60}{Fe} CR density. The triangle markers represent the median (50$^{\rm th}$ percentile) of the distribution while the lower (upper) bound of the error bar corresponds to the 16$^{\rm th}$ (84$^{\rm th}$) percentile, defining a 68\% coverage probability. We see that most of the Sco-Cen realizations host a young supernova with an age smaller than 300~kyr, except for two cases for which the explosion time distribution of supernova extends to higher values. This arises from different effects depending on the case. For smaller values of the CR diffusion coefficient, as discussed previously, one expects the age of the supernovae to be greater and to span a larger range since the propagation time for CRs is longer. In the case where the CR acceleration efficiency is increased, older supernovae, which would have contributed negligibly otherwise, may now contribute significantly to the observed \nuc{60}{Fe} CR density budget. Interestingly, the mass of the Sco-Cen ccSN accelerating more than 50\% of the observed \nuc{60}{Fe} CR density is nearly independent of the different test cases that we explored: the median value is $\approx 15-16~M_\odot$ and the $1\sigma$ range $\approx 13-20~M_\odot$.

\subsection{Geminga}
The Geminga pulsar is currently located in the constellation Gemini at a distance of about $157^{+59}_{-34}$~pc with a proper motion of $170\pm4$~mas yr$^{-1}$ \citep{Caraveo1996}. Its spin-down age deduced from the pulsar period and period derivative~\citep{Bignami1992} is 342~kyr, which can be considered as representative of its true age~\citep{Pellizza2005}. Even if these properties are well established, the place of birth of Geminga is not clearly identified yet. By tracing back the space motion of Geminga, \cite{Pellizza2005} find that Geminga was born at $90-240$~pc from the Sun, most probably inside the Orion OB1a association or the Cas-Tau OB association. Moreover, these authors conclude that the Geminga progenitor mass should not be greater than 15~$M_\odot$.

One of the main conclusion of the present study is that the Sco-Cen OB association plays a specific role in explaining the observed LISM \nuc{60}{Fe} CR density. However, our results also show $(i)$ that the Cas-Tau OB association is able to reproduce the observations, even though this is much more unlikely, and $(ii)$ that the occurrence of Orion OB1 as the most contributing association can be significant when the \cite{Sukhbold2016} explodability criterion is used (see Fig.~\ref{f:gcrRelativeContributionSukhbold}). 

Here, we investigate whether the Geminga progenitor could be the supernova that accelerated the \nuc{60}{Fe} nuclei observed by ACE/CRIS, and whether this supernova could be associated to the Orion OB1 or Cas-Tau associations. 

In the following, we consider Orion OB1 and Cas-Tau as two independent OB associations. We first compute the nucleosynthetic activity as a function of time of each OB association, and we estimate, for each realization of our model ($N=4000$), the amount of \nuc{60}{Fe} present in the associated superbubble gas when the Geminga progenitor exploded 342~kyr ago. In a second step, and for each realization, the distance of Geminga to the solar sytem $d$ is uniformly sampled up to the distance of the considered OB associations. CRs are then accelerated and propagated across the distance $d$ during a time corresponding to the age of the Geminga pulsar, and their density is calculated using Eq.~\ref{e:diffusion}. 

In the case of the Orion OB1 association we find that in $\approx6.5\%$ of the realizations (\cite{Sukhbold2016} explodability criterion) a ccSN exploding 342~kyr ago at a distance between 90 and 240~pc from the Sun is able to accelerate the 15 \nuc{60}{Fe} nuclei detected by ACE/CRIS. On the contrary, we find that for the Cas-Tau OB association on its own no supernovae is able to reproduce the ACE/CRIS observations (\cite{LC18} explodablity criterion). From these results the Geminga progenitor could be the supernova which accelerated the \nuc{60}{Fe} observed in the LISM if and only if it is associated to the Orion OB1 association. For these cases the \nuc{56}{Fe} CR density is at least ten times lower than the density measured in the LISM.

\subsection{Locality of \nuc{56}{Fe} CRs} 
\label{sec:locality}

An intriguing result of our work is that a substantial fraction of the $^{56}$Fe in the CR composition is found to be of local origin: of the order of $\sim 20$ \% in most cases and up to $\sim 50$ \% for special input parameters~(see Fig.~\ref{f:sensitivity}). 

This fact can be understood by computing the maximum distance that a CR nucleus of \nuc{56}{Fe} can diffuse in the ISM before undergoing catastrophic losses (spallation).
The characteristic spallation time for \nuc{56}{Fe} isotopes of energy equal to 550 MeV/nucleon is $\tau_{\rm spal} \sim 16~(n_{\rm H}/0.1~{\rm cm}^{-3})^{-1}$~Myr, which translates into a maximum diffusion distance of:
\begin{eqnarray}
d_{\rm max} &=& \left( 6 ~D ~\tau_{\rm spal} \right)^{1/2} \\
&\sim& 3.6 \left( \frac{D}{4 \times 10^{28} {\rm cm^2/s}} \right)^{1/2} \left( \frac{\rm n_H}{0.1~{\rm cm^{-3}}} \right)^{-1/2} \rm kpc \nonumber 
\end{eqnarray}
where the diffusion coefficient has been normalised to the appropriate value for \nuc{56}{Fe} CRs, and the gas density $n_{\rm H}$ correspond to the effective density experienced by CRs during their trip to the solar system, while propagating both through the halo and disc (for sources located at a distance larger than the thickness of the gaseous disk, CRs spend a sizeable fraction of the propagation time in the halo). This means that the \nuc{56}{Fe} CRs that we observe at the Earth have been produced at sources located at a distance smaller than $d_{\rm max}$.

More quantitatively, let's consider a situation where CRs are produced at a constant (in both space and time) rate at any location on the Galactic disk.
Let $q_{\rm CR}$ be the rate at which CRs of a given specie and of a given energy are produced within an infinitesimal surface of the disk ${\rm d} \sigma$.
Then, an observer at the Earth would measure a density of CRs coming from a region ${\rm d} \sigma$ located at a distance $R$ equal to ${\rm d} n_{\rm CR} = q_{\rm CR} {\rm d} \sigma / 4 \pi D R$ if $R < d_{\rm max}$ and ${\rm d} n_{\rm CR} \sim 0$ otherwise.
Integrating over the entire surface of the disk one can see that the local density of CRs produced within a distance $R$ scales as $n_{\rm CR} (< R) = q_{\rm CR} R/2 D$.
Therefore, the fraction $\eta$ of observed CRs coming from within a distance $R$ is simply:
\begin{eqnarray}
\label{eq:local}
&\eta& (< R) \sim \left( \frac{R}{d_{\rm max}} \right) \\
&\sim& 0.28 \left( \frac{R}{{\rm kpc}} \right) \left( \frac{D}{4 \times 10^{28} {\rm cm^2/s}} \right)^{-1/2} \left( \frac{\rm n_H}{0.1~{\rm cm^{-3}}} \right)^{1/2} \nonumber 
\end{eqnarray}
This is true provided that $d_{\rm max}$ is smaller than the size of the CR halo $H$, otherwise $\eta \sim R/H$.

Equation~\ref{eq:local} shows that a fraction of about 30\% of the observed \nuc{56}{Fe} is indeed expected to be produced in the LISM, where the star clusters listed in Table~\ref{t:ob} are located. In fact, CRs are not injected at a constant rate at any location within the disk, but are rather associated to supernova explosions. The discreteness and stochasticity of stellar explosions plays a crucial role, and results in inhomogeneities in the spatial and temporal distribution of low energy CRs \citep{phan2021}. Therefore, the result obtained by means of Equation~\ref{eq:local} should be considered only as an indicative estimate.

\section{Summary and Conclusions}
Live \nuc{60}{Fe} CRs have been detected in near-Earth space by the ACE/CRIS instrument over 17 years of operation \citep{binns2016}. The \nuc{60}{Fe} radioactive lifetime of 3.8~Myr is sufficiently long such that an origin from a nearby nucleosynthesis site is plausible, and short enough so that the nucleosynthesis sites far out in the Galaxy are plausibly beyond reach for \nuc{60}{Fe} surviving such a journey. In this paper, we thus investigated the possible local sources which may have accelerated the observed \nuc{60}{Fe} nuclei.

We developed a bottom-up model computing the CR flux at the solar sytem where the nucleosynthetic output from a massive-star group is coupled to a CR transport model. The population synthesis part of our model relies on the yields from stars and supernovae, which are properly weighted by an initial mass function using a Monte Carlo approach, addressing statistical fluctuations of stellar and star group parameters. The time profile of any nuclide abundance has thus been obtained in the gas of the superbubble which is excavated by the massive-star cluster activity. We find that among the different ingredients of the population synthesis model the explodability criterion, which determines whether a massive star ends its life as a supernova or avoids explosion, has the largest impact on the nuclide abundance in the superbubble.

Once the superbubble content in \nuc{60}{Fe} is evaluated, we determine the fraction ending up in locally-accelerated CRs, and propagate these from their source through the ISM toward the solar sytem. We consider a simple acceleration and propagation model where the advection and ionization energy losses can be neglected, and where accelerated ions, when escaping from their source, diffuse isotropically in the ISM and suffer both catastrophic and radioactive losses. Both the CR acceleration efficiency and diffusion coefficient are very uncertain, in part because of the structure of magnetic field and the superbubble environment (diffusion coefficient), and the efficiency of dust production and its destruction by thermal sputtering (acceleration efficiency).

When applying our CR population synthesis and transport model to all the OB associations within 1~kpc of our solar sytem \citep{Wright2020} we find that the 15 nuclei of \nuc{60}{Fe} detected by the ACE/CRIS instrument most probably originate from the Sco-Cen OB association. Moreover, we find that a young supernova (age $\leq 500$~kyr) with a progenitor mass of $\approx13-20~M_\odot$ might be the source of acceleration of the observed \nuc{60}{Fe} nuclei. These results are largely independent of the assumed input parameters of our model except for the explodability criterion. When the \cite{Sukhbold2016} criterion is used, the Orion OB1 association may also contribute significantly to the observed \nuc{60}{Fe} CR density in the LISM.

The Orion OB1 association and the Cas-Tau OB association are both possible birthplaces of the Geminga pulsar \citep{Pellizza2005}. We investigate  the possibility that the observed \nuc{60}{Fe} nuclei were accelerated by the SN explosion that gave birth to the Geminga pulsar, and we show that a ccSN exploding 342~kyr ago (age of Geminga) at a distance between 90 and 240~pc from the Sun (presumed distance of Geminga at its birth) can account for the observed \nuc{60}{Fe} CR density in the LSIM if, and only if, the progenitor of Geminga is located in the Orion OB1 association. The associated probability for such a case is of about $6-7\%$.

The origin of the live \nuc{60}{Fe} nuclei detected by the ACE/CRIS instrument could be traced back to the closest nearby OB associations. With the same formalism we computed the CR density of radioactive \nuc{26}{Al} and stable \nuc{56}{Fe} nuclei in the LISM. We find that the \nuc{26}{Al} density calculated from local OB associations is more than an order of magnitude lower than that deduced from ACR/CRIS observations, which confirms that \nuc{26}{Al} in CRs is mainly a secondary species produced by spallation of heavier nuclei (mainly \nuc{28}{Si}). However, we also find that about 20\% of the observed \nuc{56}{Fe} density can be accounted for by local OB associations located at less than $\sim1$~kpc from the solar system. These results are independent of the population synthesis parameters (IMF, yields and explodability), but do show a sensitivity to the CR acceleration efficiency and diffusion coefficient. Varying by a factor of 10 down and up the CR acceleration efficiency and the diffusion coefficient, respectively, the \nuc{56}{Fe} density calculated from local OB associations can represent up to 50\% of the observed value. Overall, the calculated contribution of local sources to the \nuc{56}{Fe} CR population appears to be consistent with a simple estimate assuming homogeneous CR production at a constant rate across the Galactic disc. 

\section*{Acknowledgements}
SG acknowledges support from Agence Nationale de la Recherche (grant ANR-21-CE31-0028).

\section*{Data Availability}
The stellar yields used in this work are publicly available from the works of \cite{LC18} and \cite{ebinger2019}. Data arising from the present work are available on reasonable request to the corresponding author. 



\bibliographystyle{mnras}
\bibliography{astro} 








\bsp	
\label{lastpage}
\end{document}